\documentstyle[epsf,times,maps,natbib]{jfm}

\newcommand{\ba}{{\mathbf{a}}}
\newcommand{\br}{{\mathbf{r}}}
\newcommand{\bu}{{\mathbf{u}}}
\newcommand{\dnn}{{D_{\mathrm{NN}}}}
\newcommand{\dll}{{D_{\mathrm{LL}}}}

\newcommand{\Rl}{{R_\lambda}}
\newcommand{\rhoeff}{{\rho_{\mathrm eff}}}
\newcommand{\bfw}{5in}
\newcommand{\fw}{4in}
\newcommand{\sfw}{3in}

\newcommand{\smallfigurewidth}{2.5in}
\newcommand{\tinyfigurewidth}{1.5in}
\newcommand{\td}{one-dimensional\,}
\newcommand{\ttd}{two-dimensional\,}
\newcommand{\tttd}{three-dimensional\,}

\title{Measurement of Particle Accelerations in Fully Developed Turbulence}
\author{Greg~A.~Voth, A.~La~Porta, Alice~M.~Crawford, Jim~Alexander,
Eberhard~Bodenschatz}
\affiliation{Laboratory of Atomic and Solid State Physics, Laboratory of
Nuclear Studies, Cornell University, Ithaca, NY 14853}

\begin{document}

\mmaketitle

\begin{abstract}
We use silicon strip detectors  (originally developed for
the CLEO~III high energy particle physics experiment) to measure
fluid particle trajectories in turbulence with temporal resolution
of up to 70,000 frames per second. This high frame rate allows the
Kolmogorov time scale of a  turbulent water flow to
be fully resolved for $140 \ge R_\lambda \ge 970$. Particle
trajectories exhibiting accelerations up to 16,000~$\mathrm{m} \,
\mathrm{s}^{-2}$ (40 times the rms value) are routinely observed.
The probability density function of the acceleration is found to
have Reynolds number dependent stretched exponential tails. The
moments of the acceleration distribution are calculated. The
scaling of the acceleration component variance with the energy
dissipation is found to be consistent with the results for low
Reynolds number direct numerical simulations, and with the K41
based Heisenberg-Yaglom prediction for $R_\lambda \ge 500$. The
acceleration flatness is found to increase with Reynolds number,
and to exceed 60 at $R_\lambda = 970$. The coupling of the
acceleration to the large scale anisotropy is found to be large at
low Reynolds number and to decrease as the Reynolds number
increases, but to persist at all Reynolds numbers measured. The
dependence of the acceleration variance on the size and density of
the tracer particles is measured. The autocorrelation function of
an acceleration component is measured, and is found to scale with
the Kolmogorov time $\tau_\eta$.
\end{abstract}

\date{\today}

\section{Introduction}

Fluid turbulence may be characterized in terms of
variables defined at points fixed in space (the Eulerian reference frame),
or in terms of the trajectories of fluid particles
(the Lagrangian reference frame).
This distinction applies both to theoretical formulations of
turbulence and to experimental techniques for
characterizing turbulent flows.
Although the formulation of fluid dynamics is
generally considered to be more tractable
in terms of Eulerian variables,
the critical issues of transport and mixing in turbulence
are more directly related to the properties of fluid
trajectories \citep{shraiman:2000,yeung:review,sawford:2001},
and are often addressed
using Lagrangian techniques \citep{pope:1994}.
There are also many applications in which the transport or
aggregation of particulate matter in turbulence is important
in its own right, such as water droplet aggregation
in clouds \citep{vaillancourt:2000} or the industrial
production of nanoparticles \citep{pratsinis:1996}.

In a basic sense, data obtained from Lagrangian
and Eulerian measurements are complementary.
In the Eulerian frame, one is typically concerned
with differences between quantities (velocity component, scalar concentration,
etc.) measured for several points separated by a fixed distance in space.
It would be equally interesting to study \textit{time} differences,
however such measurements made within the Eulerian framework are
difficult to interpret, since the large scales of the flow will
sweep the turbulence past a fixed detector, causing temporal structure
to be entangled with spatial structure.
The true temporal structure of turbulence is only revealed when fluctuations
are measured along a particle trajectory, in the Lagrangian frame.

Turbulence has traditionally been studied in the Eulerian frame
rather than the Lagrangian frame for technical reasons. The hot
wire anemometer used in conjunction with the Taylor frozen flow
hypothesis provides extremely accurate Eulerian data in turbulent
gas flows over a broad range of Reynolds number, but no comparably
effective technique has been available for Lagrangian
measurements. However, as the range of Reynolds number accessible
to direct numerical simulations (DNS) of turbulence has expanded,
numerical studies of particle trajectories in isotropic turbulence
have yielded important insights
\citep{vedula:1998,yeung:2001,gotoh:1999,gotoh:2001}. 
Although very little experimental data
for Lagrangian properties in fully developed turbulence has been available,
this is beginning to change. 
Basic issues, such as the Richardson law for
particle dispersion \citep{Monin:1975:SFM},  the
Heisenberg-Yaglom prediction of fluid particle accelerations
\citep{heisenberg:1948,yaglom:1949}, and Kolmogorov scaling of
temporal velocity differences  have remained untested 
 for many decades. Recent experiments  by 
\cite{ott:2000}, \cite{Voth:1998:LAM}, and \cite{pinton:2001} have made significant progress in
addressing each of these but there are still major limitations. In particular,
limitations in spatial and temporal measurement resolution have obscured the small scales at
large Reynolds number. More complete information 
about particle trajectories in real (non-idealized) turbulent
flows is needed to guide the development of models of transport in
applications \citep{pope:1994}. 

In principle, fluid particle trajectories are easily measured by
seeding a turbulent flow with minute tracer particles and following
their \tttd motions with an imaging system.
In practice, this is a very challenging task because changes
in particle velocity or acceleration can
take place on time scales of order of the Kolmogorov time,
$\tau_\eta = (\nu/\epsilon)^{1/2}$ where $\nu$ is the kinematic viscosity
and $\epsilon$ is the energy dissipation per unit mass.
In order to observe universal scaling behaviour we require that
the Reynolds number, defined by
${\mathrm Re} = u L/\nu$, where $u$ is the rms velocity and $L$ is the
energy injection scale, approach $10^5$.
(This is equivalent to requiring that the Taylor microscale Reynolds
number $\Rl = (15 \mathrm{Re})^{1/2}$ approach 1000.)
In a laboratory water flow
($\nu \approx 10^{-6}\ {\mathrm m}^2\,{\mathrm s}^{-3}$)
with convenient energy injection
scale ($L \approx 0.1\ {\mathrm m}$) and assuming $\epsilon = u^3/L$
we must have $\epsilon \approx 10\ {\mathrm m}^2\,{\mathrm s}^{-3}$,
which implies $\tau \approx 0.3\ {\mathrm ms}$.
In order to measure particle accelerations, motions which take
place over this time scale must be fully resolved.

Although conventional imaging systems based on charge coupled
devices (CCD) have been
used for \tttd tracking of particles in low Reynolds number
flows \citep{Snyder:1971:SMP,Sato:1987:LMF,virant:1997,ott:2000}
they do not provide adequate temporal resolution for use
in fully developed turbulence, as specified above.
However, these requirements may be met by the use of silicon
strip detectors as optical imaging elements in a particle
tracking system.
The strip detectors used in our experiment were
developed to measure sub-atomic particle tracks in the vertex detector
of the CLEO~III experiment operating at the Cornell Electron-Positron
Collider ~\citep{skubic:1998}.  When applied to particle tracking in turbulence
each detector measures a one dimensional projection of the
image of the tracer particles.
Using a data acquisition system designed specifically
for the turbulence experiment
several detectors may be simultaneously read out at 70,000
frames per second, making it possible to measure
\ttd or \tttd particle trajectories with very high
spatial and time resolution \citep{voth:2001, laporta:2001}.


\begin{figure}
\begin{center}
\epsfxsize=\smallfigurewidth
\mbox{%
\epsffile{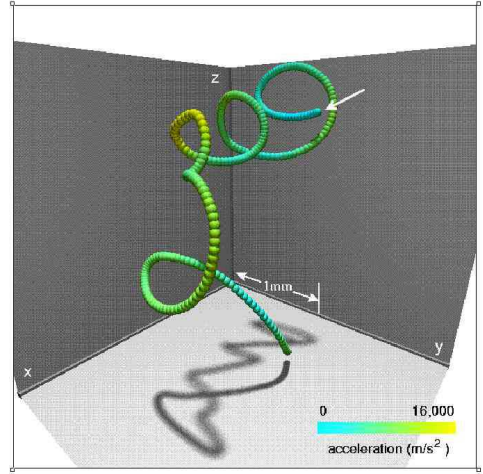}%
}
\end{center}
\caption{Trajectory of a 46~$\mu$m diameter tracer particle in
turbulence at $\Rl = 970$ recorded at frame rate 70,000~fps.
The position of the particle
at each of 278 frames is represented as a sphere.  The acceleration
magnitude is represented as the colour of the trajectory, as indicated
by the scale.}
\label{fig:traj}
\end{figure}

\begin{figure}
\begin{center}
\epsfxsize=\fw
\mbox{%
\epsffile{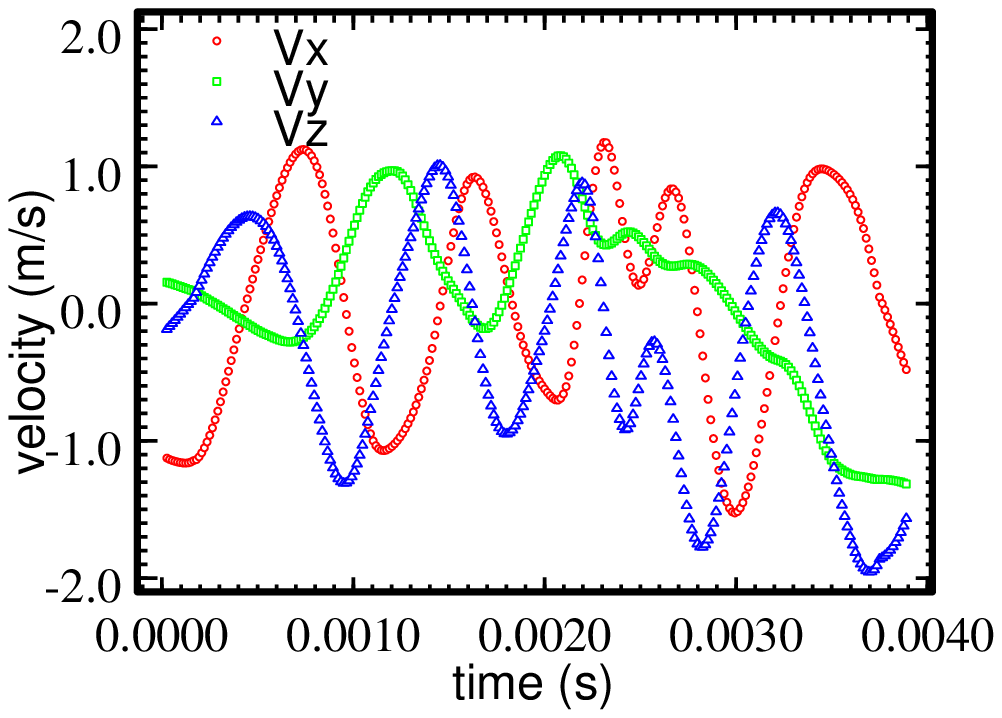}%
}
\end{center}
\caption{Three components of the velocity of the particle
shown in figure~\ref{fig:traj_vel}.
}
\label{fig:traj_vel}
\end{figure}

The technical demands of particle tracking in turbulence and the
scope of phenomena which become accessible when these demands are
met can best be appreciated by examining one of the more dramatic
particle trajectories recorded with the strip detector
particle tracking system.
The \tttd trajectory shown in figure~\ref{fig:traj} was recorded
in a flow between counter-rotating disks at $R_\lambda = 970$
(described below in Section~\ref{sec:flow})
having an rms velocity of approximately 1~$\mathrm{m}\,\mathrm{s}^{-1}$,
a Kolmogorov time $\tau_\eta$ of approximately 1/3~ms, and a
Kolmogorov distance scale  $\eta$ of 20~$\mu$m.

The trajectory was recorded
at 70,000 frames per second using a pair of strip detectors
collecting primary and conjugate charge,
as will be described in Section~\ref{sec:experimental} below.
The particle enters the measurement volume near the top right of
figure~\ref{fig:traj} and appears to be trapped in a vortical structure.
In the third, tightest turn of the helical motion the acceleration
of the particle rises to 16,000~$\mathrm{m}\,\mathrm{s}^{-2}$ within
a time interval of 0.5~ms ($\approx 1.5 \tau_\eta$).
During this trajectory the velocity
components of the particle, shown in figure~\ref{fig:traj_vel}, oscillate
wildly, spanning a range considerably
larger than the rms velocity within a Kolmogorov time.
The extreme fluctuations in velocity and acceleration,
which occur in time-scales of
order of the Kolmogorov time, make great demands on the particle
tracking system.

The trajectory in figure~\ref{fig:traj} and \ref{fig:traj_vel} is also
noteworthy in that it seems to lie outside the characteristic
range of the energy cascade.
For the first part of the trajectory the particle
appears to be caught an a very intense vortical structure.
The tightest loop in the helical trajectory of the particle
is about 300~$\mu$m in diameter, or about 15~$\eta$, which is
in the inertial subrange quite close to
the dissipative scale of the turbulence.
The period of the motion
appears to be less than 1~ms, of the
order of the Kolmogorov time $\tau_\eta$.
Yet the velocity fluctuations observed in this trajectory exceed the
rms velocity, which would normally be associated with the
largest scales of the turbulence (the so called energy-containing range).
Events of this nature, which up until now were experimentally inaccessible,
may be interpreted as a manifestation of turbulent intermittency.

The remainder of the paper is organized as follows.
In Section~\ref{sec:experimental} the operating principles and
capabilities of the silicon strip detector based particle
tracking system are discussed.
This includes discussion of the optical system used to image the
tracer particles on the strip detector (Section~\ref{sec:optics}).
In Section~\ref{sec:flow} the turbulent water flow between
counter-rotating disks is described, and characterized using
the silicon strip detector.
This entails measurement of the velocity statistics and estimate
of the energy dissipation from the transverse velocity structure
function.

The main results are the investigation of the statistics of
particle accelerations in turbulence for $140 \le \Rl \le 970$,
given in Section~\ref{sec:results}.  (A brief account of part of
this research has previously been published in \textit{Nature}
\citep{laporta:2001}.) In Section~\ref{sec:apdf} the probability
density function (PDF) of the acceleration component is measured
as a function of Reynolds number and component direction. In
Section~\ref{sec:avariance} the scaling of the acceleration
component variance is compared with the Heisenberg-Yaglom
prediction. It is found that data is consistent with DNS results
at low Reynolds number, and with the Heisenberg-Yaglom predicted
scaling for $R_\lambda \ge 500$. The flatness of the acceleration
component is found to be quite large and to increase with Reynolds
number, exceeding 60 at $R_\lambda = 970$. The acceleration
component autocorrelation function is shown in
Section~\ref{sec:acorrelations}. The autocorrelation function is
shown to scale with the Kolmogorov time, and to cross zero at a
time of approximately $2.1~\tau_\eta$. Finally, the influence of
particle size and density are studied in Section~\ref{sec:psize}.
It is found that the acceleration of relatively large particles
(diameter $\ge$ 5 $\tau_\eta$) can be significantly smaller than
that of infinitesimal fluid particles, and that, in this
experiment, particle size and not density is primarily responsible
for this. However, the 46~$\mu$m particles used for the
acceleration measurements differ from ideal fluid particles by
only a few percent. Two appendices are also included. The first
details the algorithms used for extracting particle tracks from
the strip detector output, and the second compares the results
published here with previous measurements, which were made with a
conventional detection system which had much lower resolution.

\section{Particle Tracking System}
\label{sec:experimental}

\subsection{Strip Detector}
\label{sec:stripdetector}

\begin{figure}
\begin{center}
\epsfxsize= \fw
\mbox{%
\epsffile{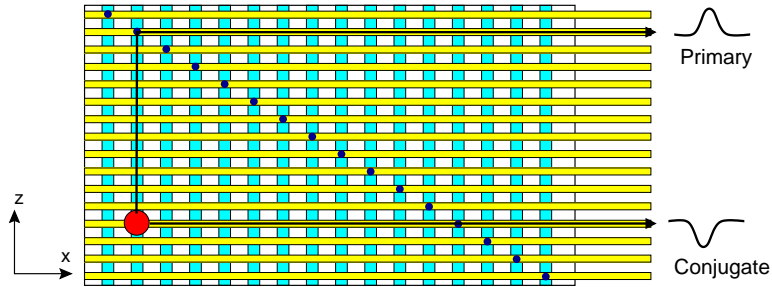}%
}
\end{center}
\caption{Simplified schematic of the strip detector.
Dark grey bars represent p-type sense strips.  Light grey
bars superimposed on the dark bars
represent metal leads evaporated on to the surface of the detector.
The small circles along the diagonal represent paths that
connect sense strips to the leads.  Under normal circumstances,
only ``primary'' charge is collected indicating the $x$ coordinate
of light spots incident on the detector.  However, the detector bias may
be adjusted so that ``conjugate'' charge is also detected by the metal
leads indicating the $z$ coordinate, as described in the text. }
\label{fig:stripd}
\end{figure}

\begin{figure}
\begin{center}
\epsfxsize= \fw
\mbox{%
\epsffile{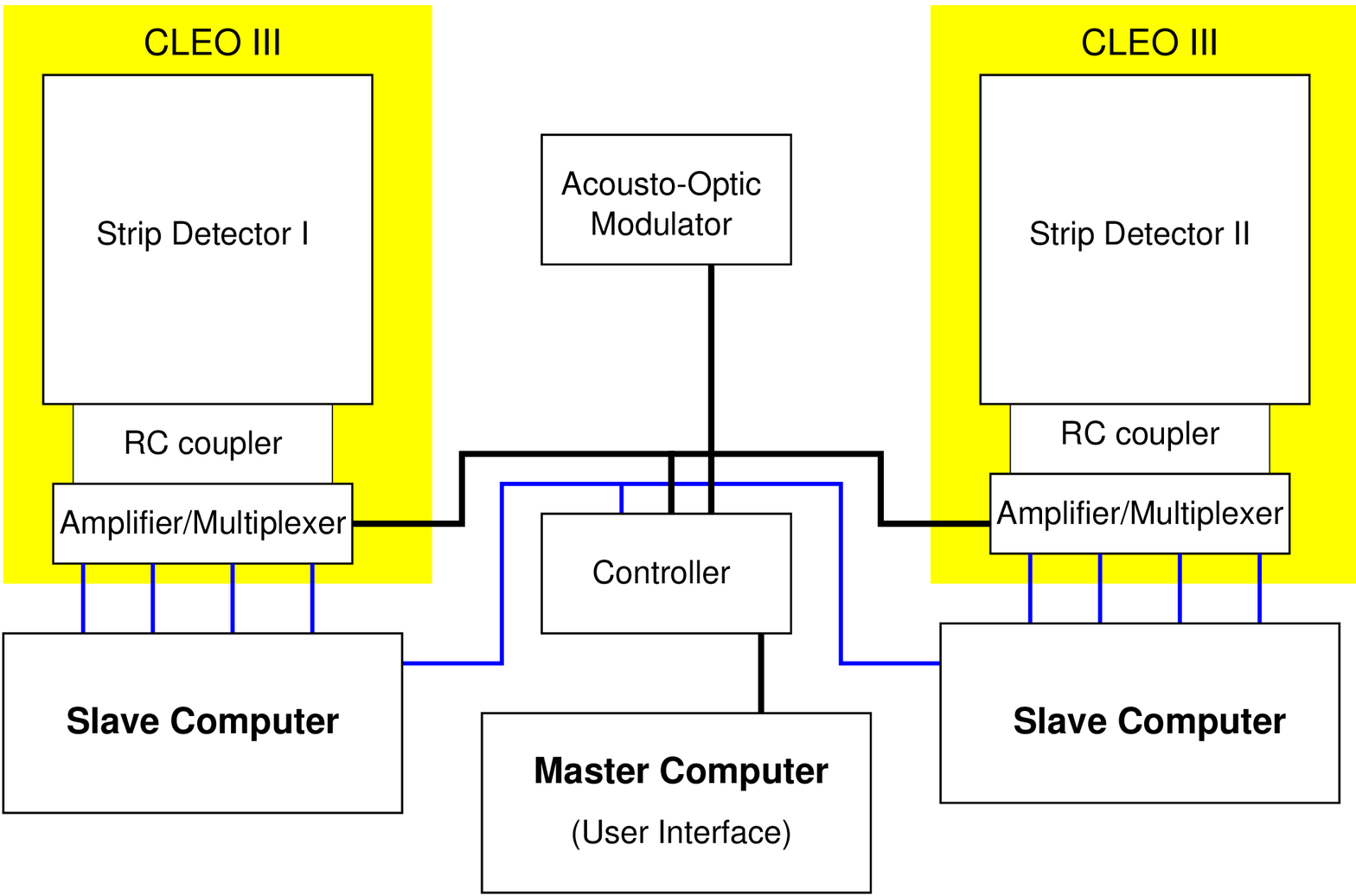}%
}
\end{center}
\caption{
The readout cluster.
Each strip detector is read out using an individual slave
computer.
A field programmable gate array (FPGA) detector controller generates timing signals for the
strip detector data acquisition hardware, and acousto-optic modulator.
A master computer controls the slave computers via Ethernet messaging
and provides the user interface for the system.
}
\label{fig:cluster}
\end{figure}

A brief description of the the strip detector tracking system is  provided below.
It is described in detail in
 \citep{voth:2001}.

The silicon strip detector is essentially a large planar photodiode
(sensitive area 51~mm $\times$ 25.6~mm) which is
segmented into 512 sense strips, as shown in figure~\ref{fig:stripd}.
Optical radiation incident on the detector creates electron-hole
pairs, and the holes are collected by p-type strips patterned on
the front surface of the n-type detector.
Charge is conducted off the detector chip through
metallic leads evaporated onto the surface of the detector which are
oriented perpendicular to the sense strips.
The positive charge collected by the array of strips gives a
\td projection of the light intensity incident on the detector,
allowing the $x$ coordinate of a particle that is imaged onto the
detector to be measured.
However, for certain settings of the detector bias, electrons appear to
become trapped in meta-stable surface states and couple capacitively
to the metallic leads.
Under these circumstances, negative ``conjugate''
charge indicates the projection of the intensity on the axis defined by
the leads (which is perpendicular to the axis defined by the sense strips).
By distinguishing the primary (positive) and conjugate (negative) peaks the
\ttd projection of a particle trajectory
may be measured  using a single detector.
Positions measured from conjugate charge have greater uncertainty,
so for most of the data presented below, and in particular for the
acceleration measurements, primary charge was used for position measurement.

The 512 strips of each strip detector are connected to an array of four
resistor-capacitor (RC) chips which provide an individual bias resistor and coupling
capacitor for each channel.
The output capacitors are connected to an array of four
integrated amplifier/multiplexer chips, which provide a shaper
amplifier and sample-and-hold amplifier for each strip.
After a frame of data is latched on the array of sample and
hold amplifiers, analog multiplexers integrated into the amplifier
chips output the strip intensities as four 10~Msample/sec waveforms.
A detector controller generates frame-readout triggers for the strip
detectors, readout electronics and an acousto-optic modulator,
and may be configured for readout rates in the range 5~kHz--70~kHz.
The acousto-optic modulator is used to strobe the illumination before
the readout of each frame, which is necessary to optimally drive
the shaper amplifiers.
The output waveforms are captured using a pair of dual-channel digital
oscilloscope boards mounted in a PCI bus {\em slave} computer.
The oscilloscope boards can store 4,000 frames of data in internal
memory, which is subsequently
downloaded into the computer's main memory.
The slave computer performs pedestal subtraction, thresholding,
and stores compressed data to a local hard disk concurrently with
the acquisition of the next frame of data.
The maximum duration of continuous acquisition ranges from
800~ms at 5~kHz to 58~ms at 70~kHz.
The time required to process and store one 4000 frame sequence
is approximately 1~s, and this limits the duty cycle of the
acquisition system at high frame rates.

\subsection{Optical Imaging System}
\label{sec:optics}

\begin{figure}
\begin{center}
\epsfxsize= \fw
\mbox{%
\epsffile{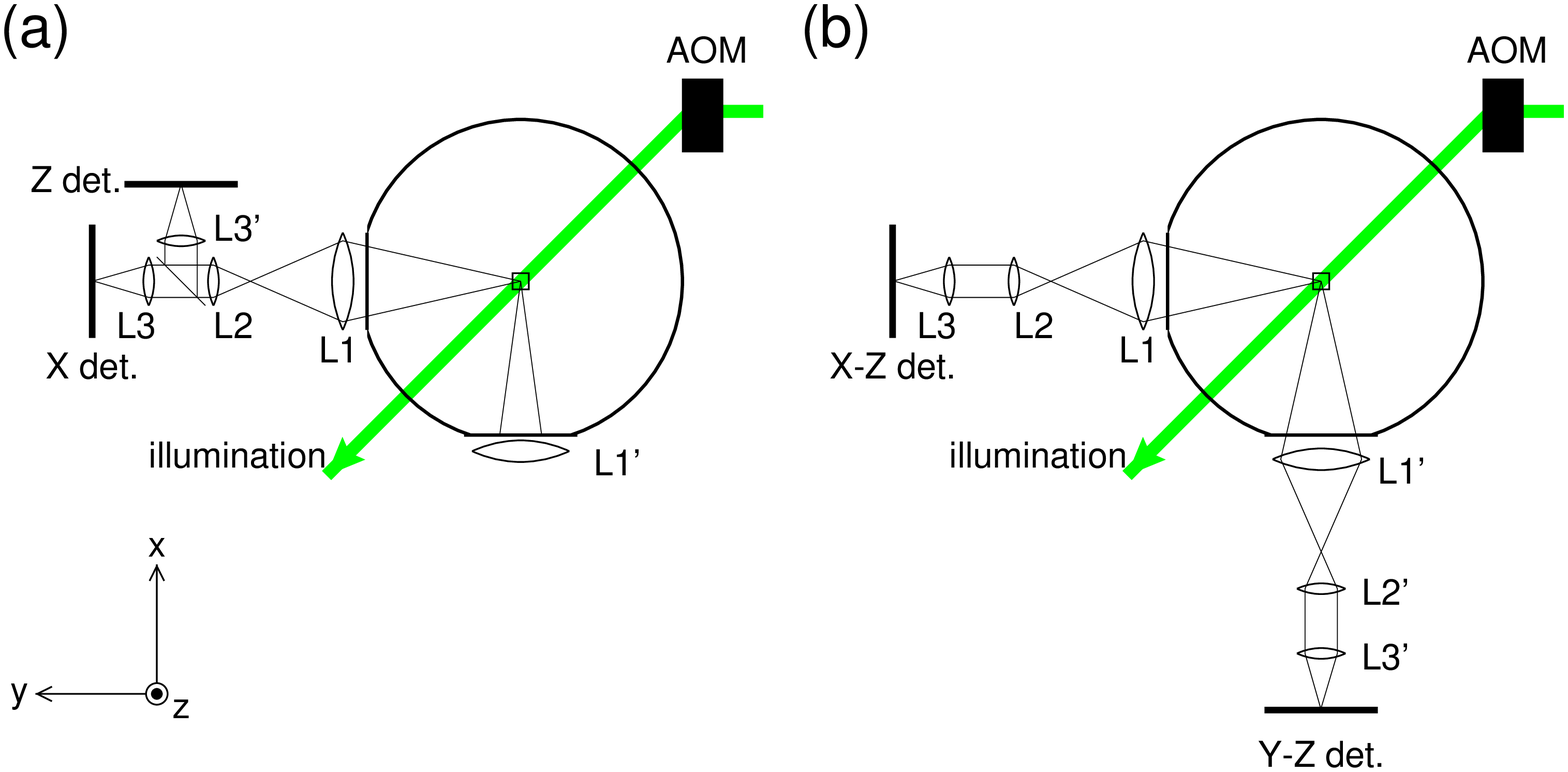}%
}
\end{center}
\caption{ (a) Optical layout for acceleration measurements, viewed
along the axis of the cylindrical turbulence chamber ($\hat{z}$).
The illumination beam is gated by an
acousto-optic modulation (AOM) before being passed through the centre
of the turbulence chamber.   Optics are used to image the central volume
through a view-port at $45^\circ$ with respect to the illumination.
A beam-splitter allows the image to be projected on two strip detectors
which are oriented to measure two orthogonal coordinates.
An additional view-port may also be used to measure the third coordinate.
(b) Optical layout for \tttd particle measurement.  The apparatus is
the same except that imaging optics are present on both ports.
In this case strip detectors are operated with conjugate peaks enabled
(see text) so each port has a detector which measures two position coordinates.
on each port. }
\label{fig:app}
\end{figure}

\begin{figure}
\begin{center}
\epsfxsize= \fw
\mbox{%
\epsffile{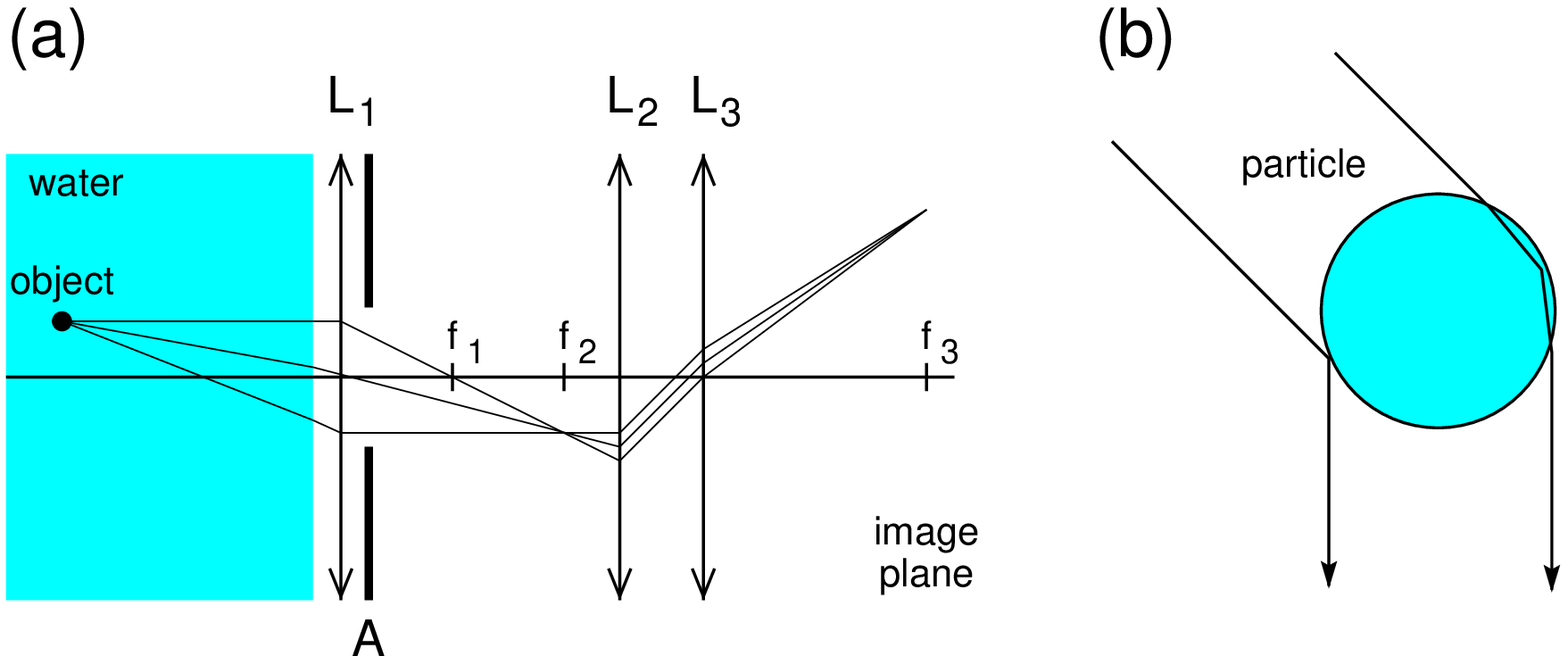}%
}
\end{center}
\caption{ A simplified schematic of the optical system is shown.
Plano-convex lens $L_1$ is place just outside the fluid chamber and
is used to image the active volume with 1:1
magnification.
A short focal length lens $L_2$ is positioned so that its front
focal plane ($f_2$) lies at
the centre of the image formed by $L_1$.
As a result, point sources in the image are transformed into parallel
ray bundles by $L_2$.
The bundles are focused by $L_3$ at its rear focal plane ($f_3$).
A beam splitter may be inserted between $L_2$ and $L_3$ to allow
the image to be projected on a second detector,
as shown in figure~\protect\ref{fig:app}.
The magnification of the system is given by the ratio of the focal
lengths $f_3/f_2$.  The light collection of the imaging system
is controlled by an aperture placed behind lens $L_1$.
(b) The tracer particles are transparent polystyrene spheres with
density 1.06~$\mathrm{g} \, \mathrm{cm}^{-3}$ and diameter 46~$\mu$m and are
detected via specular reflection from the internal and external
surfaces.  With illumination at $45^\circ$ and polarization perpendicular
to the plane of incidence, the internal reflection is much stronger than
the external reflection so that the particle appears as a single point-like
source of light.}
\label{fig:image}
\end{figure}

The optical configuration used to image particles is shown in
figure~\ref{fig:app}.
The illumination beam, generated by a 6~W continuous wave Argon-Ion laser
and gated by the acousto-optic modulator, is directed through a glass
window and passes through the centre
of the flow chamber.  The beam is a $\mathrm{TM}_{00}$ Gaussian mode and
a beam expander is used to obtain
a spot radius $\omega \approx 1.0 \mathrm{mm}$ with negligible
divergence.
This beam radius is chosen to so that the beam fills  the field of
view of the detector.
Acceleration measurements
were taken in the configuration shown in
figure~\ref{fig:app}(a), in which two strip detectors view a common
image and measure \ttd coordinates.
Some \tttd tracks were recorded in the configuration
shown in figure~\ref{fig:app}(b),
in which two detectors view the volume from different view-ports so that
their detection volumes overlap.

The design of the image system,
sketched in figure~\ref{fig:app}
is determined by the requirement that a small
measurement volume be imaged at high magnification by optics which are
placed outside the turbulence chamber.
These design requirements were met by the two stage imaging system
shown in figure~\ref{fig:image}(a).
In order to maintain acceptable depth of field,
the aperture shown after the first imaging lens
(L1 in figure~\ref{fig:image})
is used to restrict the numerical aperture of the imaging system to
a value of 0.03, giving a
spot size $\approx  300 \mu\mathrm{m}$ for particles
throughout the measurement volume.
As a result, at least three strips were illuminated by each particle
image, making it possible to locate the particle with
sub-strip resolution by fitting each peak to a Gaussian function.

Although the magnification of the optical
system defined in figure~\ref{fig:image}
is nominally $f_3/f_2$, it is useful to vary the configuration
of the system in order to adjust the magnification and
position of the focal plane.
In particular, $L_1$ may be repositioned to change the magnification of the
first stage of the system, and $L_3$ is replaced by a pair of lenses whose
separation dependent effective focal length replaces $f_3$.
For the acceleration data presented below, $L_1$ is a compensated doublet
with focal length 15~cm, $L_2$ is a doublet of focal length 3.8~cm
and $L_3$ is replaced by a pair of lenses consisting of a 30~cm focal
length plano convex lens and a $-$50~cm plano concave lens separated by
approximately 10~cm.
The detector is placed approximately 25~cm behind the plano concave lens.
The magnification of the system is approximately 12.8,
so that the 100~$\mu$m pitch
of the strip detector corresponds to approximately 7.8~$\mu$m
in the fluid volume,
giving a field of view of
2.00~mm~$\times$4.00~mm.
The effective magnification of the system is measured to within 2\%
using a
calibration target that is positioned in the active region at the centre
of the flow chamber.
(The variation of the magnification
is less than 1\% over the illuminated
volume, so it was not necessary to express the magnification
as a function of depth.)

\begin{figure}
\begin{center}
\epsfxsize=\tinyfigurewidth
\mbox{%
\epsffile{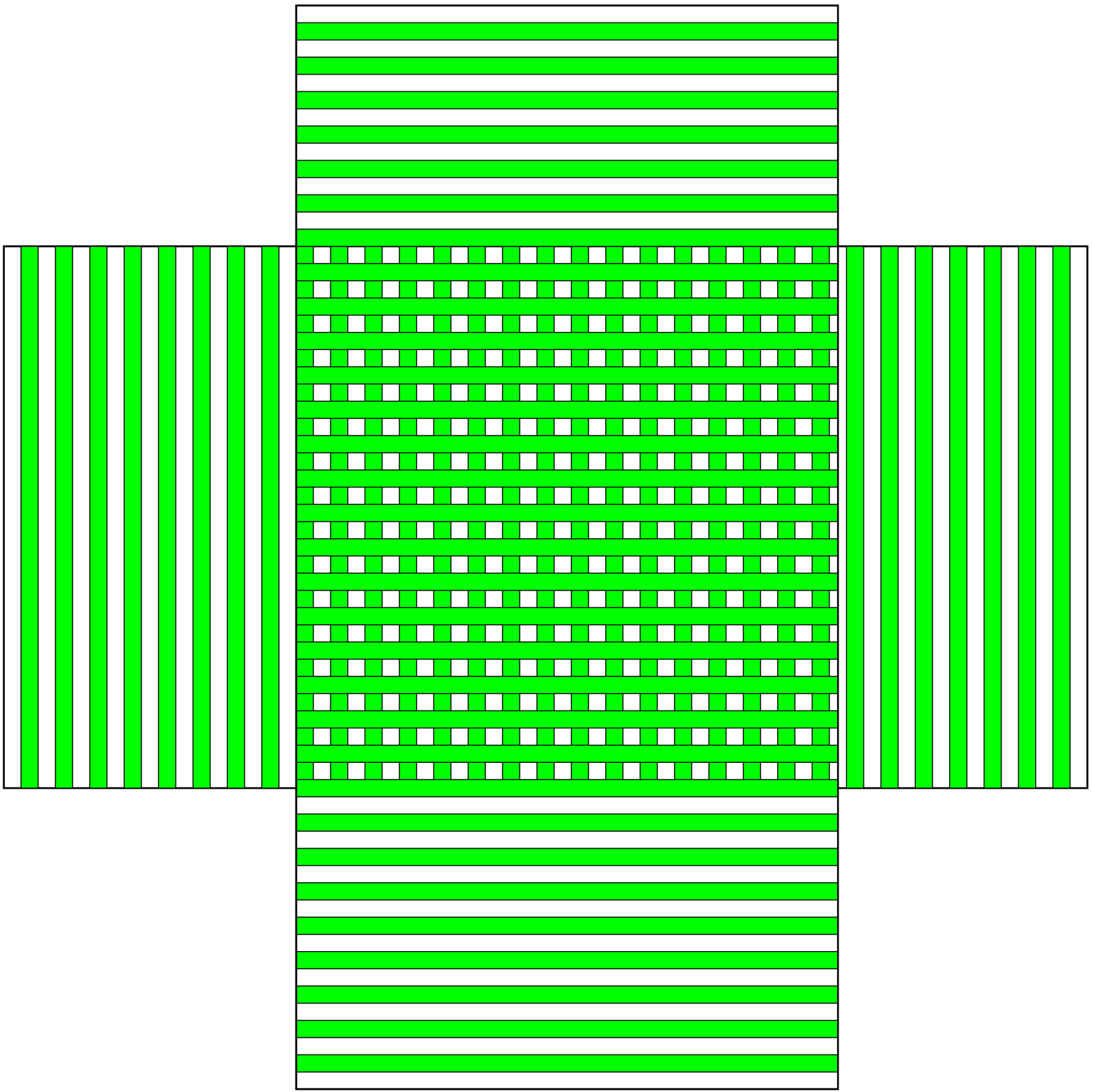}%
}
\end{center}
\caption{
The relative fields of view of two strip detectors oriented as
in figure~\ref{fig:app}(a) to measure
perpendicular coordinates, where the grey stripes represent the orientation
of the charge collecting sense strips.
Because of the 2:1 aspect ratio of the detector, only half of the field
of each detector  overlaps with the field of the other, so that only
256 strips can be used for coincident measurements.
}
\label{fig:sdpair}
\end{figure}

\begin{figure}
\begin{center}
\epsfxsize=\smallfigurewidth
\mbox{%
\epsffile{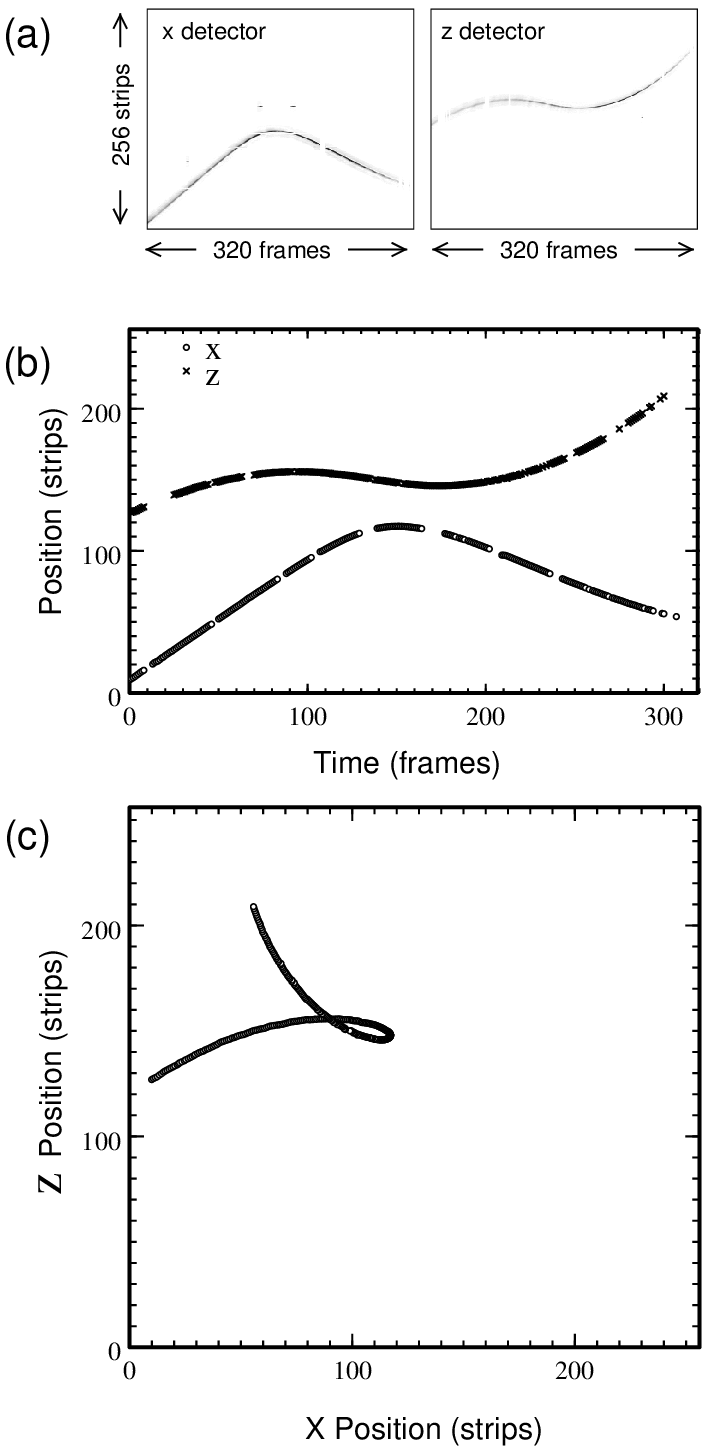}%
}
\end{center}
\caption{(a) Raw data from the two strip detectors.  Only the
256 overlapping strip segments are shown.
(b) Positions vs time measured from the raw data, where gaps are
due to the presence of inoperative strips.
(c) Reconstruction of the \ttd track from the position data in (b),
where gaps have been bridged by linear interpolation.}
\label{fig:raw_data}
\end{figure}

For acceleration data, the strip detectors are biased such
that conjugate charge is suppressed
allowing them to measure a single coordinate using the primary charge signal.
In order to measure \ttd trajectories, a beam-splitter is used
to project the same image onto two detectors (see figure~\ref{fig:app}(a)),
which are oriented
to measure orthogonal coordinates.
Because of the 2:1 aspect ratio of the detector, the fields of view
of the two detectors do not coincide, as shown in figure~\ref{fig:sdpair}.
In this case, the full area of each detector can be used to measure
\td trajectories, but \ttd trajectories can only be measured in the
region where the two fields overlap.
As a result, only half of the 512 strips are available when
analyzing \ttd trajectories.

\subsection{Determination and Characterization of Tracks}

\begin{figure}
\begin{center}
\epsfxsize=\fw
\mbox{%
\epsffile{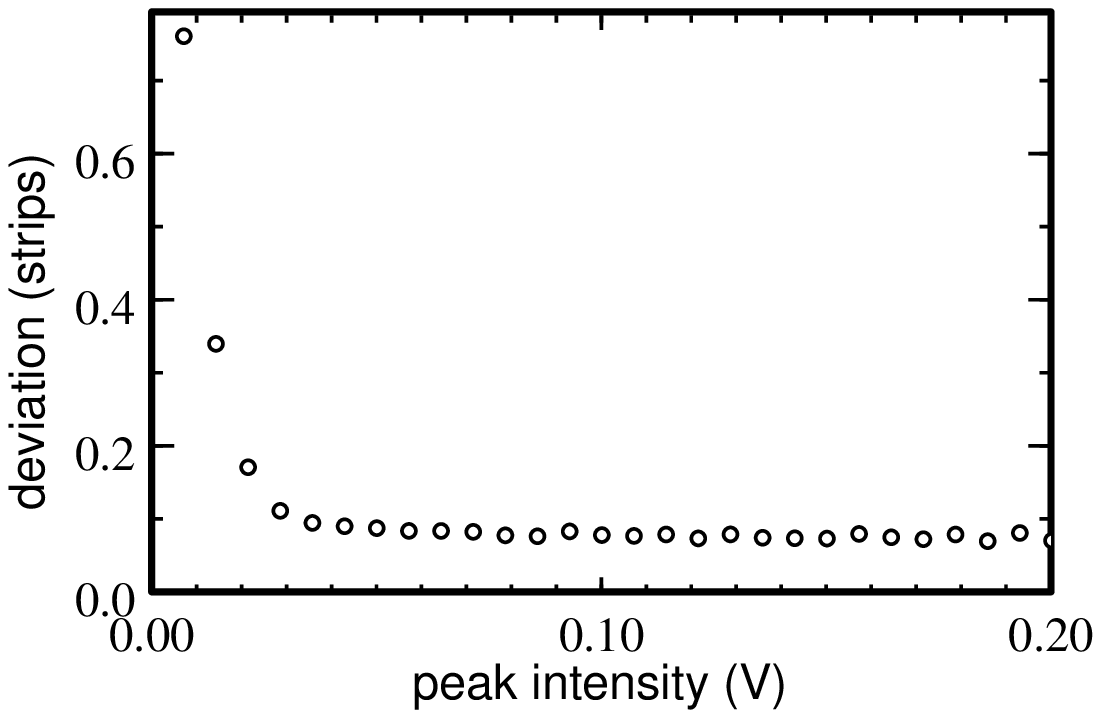}%
}
\end{center}
\caption{Estimate of particle position error.
The graph shows the rms deviation of the peak centre from a linear fit to
a short segment of track.}
\label{fig:strip_dev}
\end{figure}

\begin{figure}
\begin{center}
\epsfxsize=\fw
\mbox{%
\epsffile{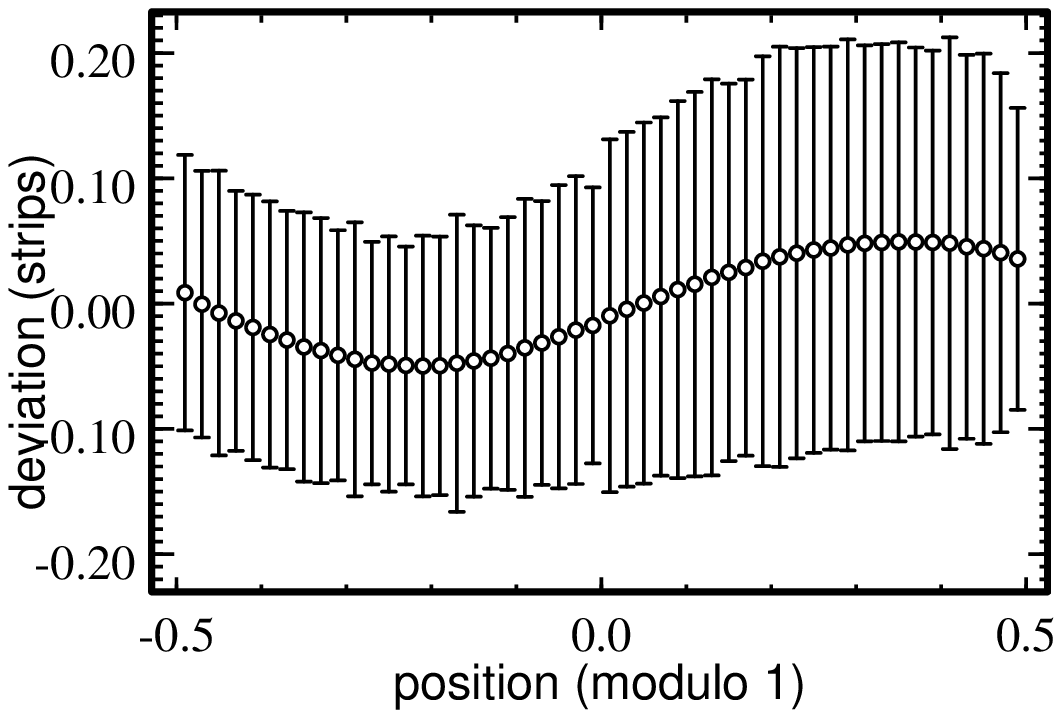}%
}
\end{center}
\caption{
Deviation of particle position from fit as a function of position
modulo 1.  The circle indicates the mean
and the error bars indicate the standard deviation of the deviation of the
peaks from the fit.}
\label{fig:dev_vs_pos}
\end{figure}

Data files stored by the experiment consist of thresholded intensity
data simultaneously acquired by the two strip detectors.
These data files are passed through several stages of analysis.
This task is similar in principal to that employed in standard
particle tracking studies using CCD cameras, but the algorithms
employed must be tailored to the peculiar characteristics
of \td projection images.
A description of the algorithm used for track extraction
is given in Appendix~\ref{sec:trackextraction}.
Secondary data files, consisting of lists of tracks measured
for each data sequence are stored for subsequent analysis.

The accuracy of the particle positions is estimated by fitting a straight
line to very short track segments (shorter than $\tau_\eta$) and measuring
the mean deviation between the data points and the fit.
The deviation is found to depend most significantly on the
peak height.
Figure~\ref{fig:strip_dev} shows the error as a function of peak
height for a typical run.
For extremely weak peaks (maximum intensity $<$ 0.02~V)
the error can be of order 1/2 strip, but the the analysis routines
are configured to limit the data
to larger intensity, so that the mean error is $\le 0.1$~strips.
The error for individual trajectories
can vary depending on focus and aperture, but always remains
below 0.2~strips.

One can also consider the mean deviation of the peak from the linear fit
as a function of the position (in strips) modulo 1, shown in
figure~\ref{fig:dev_vs_pos}.
The dependence of the mean deviation on the position indicates a
nonlinearity in the interpolation of the peak centres which is
smaller than, but of the same order as the random uncertainty.
This nonlinearity is compensated for by subtraction of the mean deviation
from the peak positions before processing of the trajectories.

The kinematic properties
of the tracks are measured by polynomial
fits (parabolic for acceleration, linear for velocity)
to the position vs time data.
The fits are made using the standard least squares
algorithm, with the relative
weight of each data point proportional to the inverse square of
the estimated error
(as a function of peak intensity, figure~\ref{fig:strip_dev}).
The length of track over which the fit is performed
is a complex issue, which is discussed in detail below.

One of the greatest challenges in the measurement of accurate
statistics of Lagrangian variables is the control of sample
biases.  From a physical standpoint, one must take care that the
tracer particles uniformly sample the fluid volume and follow the
flow field. This requires that the particles be sufficiently small
and that the density match between the particle and the fluid be
sufficiently close. This issue is addressed in
Section~\ref{sec:psize}, below. It is also necessary to ensure
that the measurement and analysis procedures do not introduce
biases in the measurements of the particle trajectories. For
instance, if one were to make one measurement for each particle
that enters the measurement volume, a distorted velocity
distribution would be obtained, since the rate at which particles
enter the measurement volume is itself proportional to the
velocity of the particle \citep{Buchave:1979:MTL,Voth:1998:LAM}.
However, the time that a particle will remain in the measurement
volume is inversely proportional to the velocity, and these two
factors cancel. Ideally, one would achieve uniform sampling within
the measurement volume by continuously measuring the kinematic
properties of the track from the time that a particle enters the
volume to the time that it exits. In practice this is impossible
to achieve because the acceleration and velocity are measured by
fitting to a polynomial function, and so the variables can not be
measured until the particle has been in the measurement volume for
a finite time. Another difficulty is that measurement is not
possible when particles cross paths or traverse inoperative
pixels. The strategy employed is to measure the variables as many
times as possible along the trajectory, and make the total
statistical weight of these measurements proportional to the total
length of the track. This seems to give the best approximation of
uniform sampling of the measurement volume.

\section{Characterization of Flow}
\label{sec:flow}

The goal of this study is to explore universal characteristics
of turbulent flows.
However, our flow deviates significantly from
the ideal of homogeneity and isotropy.
The apparatus used to generate the turbulent water flow
is described in Section~\ref{sec:gflow}.
Standard techniques for characterizing turbulence using
anemometers are not feasible.
Therefore, we have used
the strip detector particle tracking system to characterize
the flow in terms of standard Eulerian quantities.
This includes measurement of the scaling of the rms velocity
components with propeller speed
(Section~\ref{sec:velocity}),
an estimate of the rate of energy
dissipation in terms of the velocity structure functions
(Section~\ref{sec:dissipation}), and
a measurement of the anisotropy of the mean velocity field
at the centre of the flow (Section~\ref{sec:strain}).

\subsection{Turbulence Generator}
\label{sec:gflow}

\begin{figure}
\begin{center}
\epsfxsize= \fw
\mbox{%
\epsffile{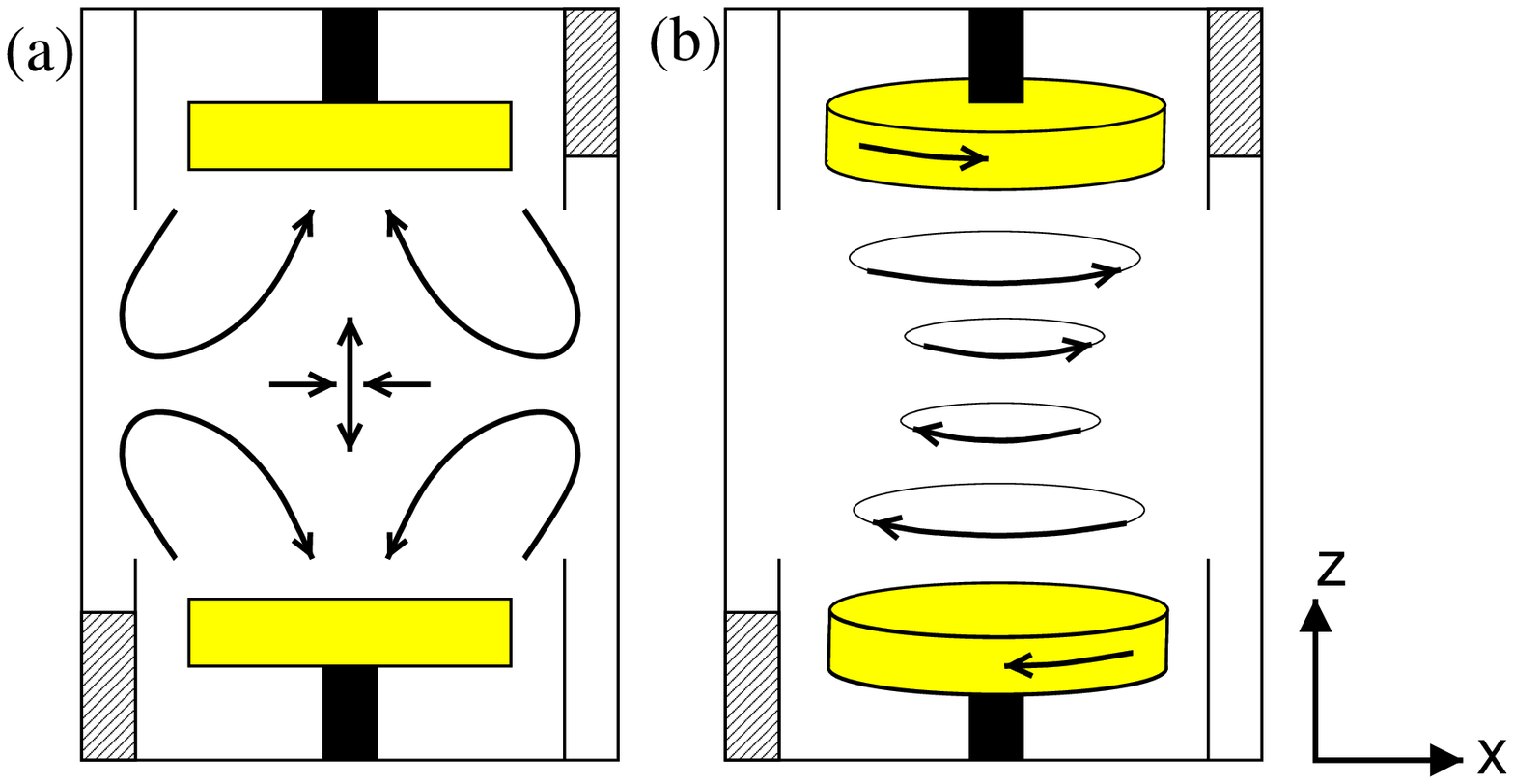}%
}
\end{center}
\caption{Schematic representation of the flow between counter-rotating
disks decomposed into (a) the pumping mode and (b) the shearing mode.}
\label{fig:flow}
\end{figure}

\begin{figure}
\begin{center}
\epsfxsize= \fw
\mbox{%
\epsffile{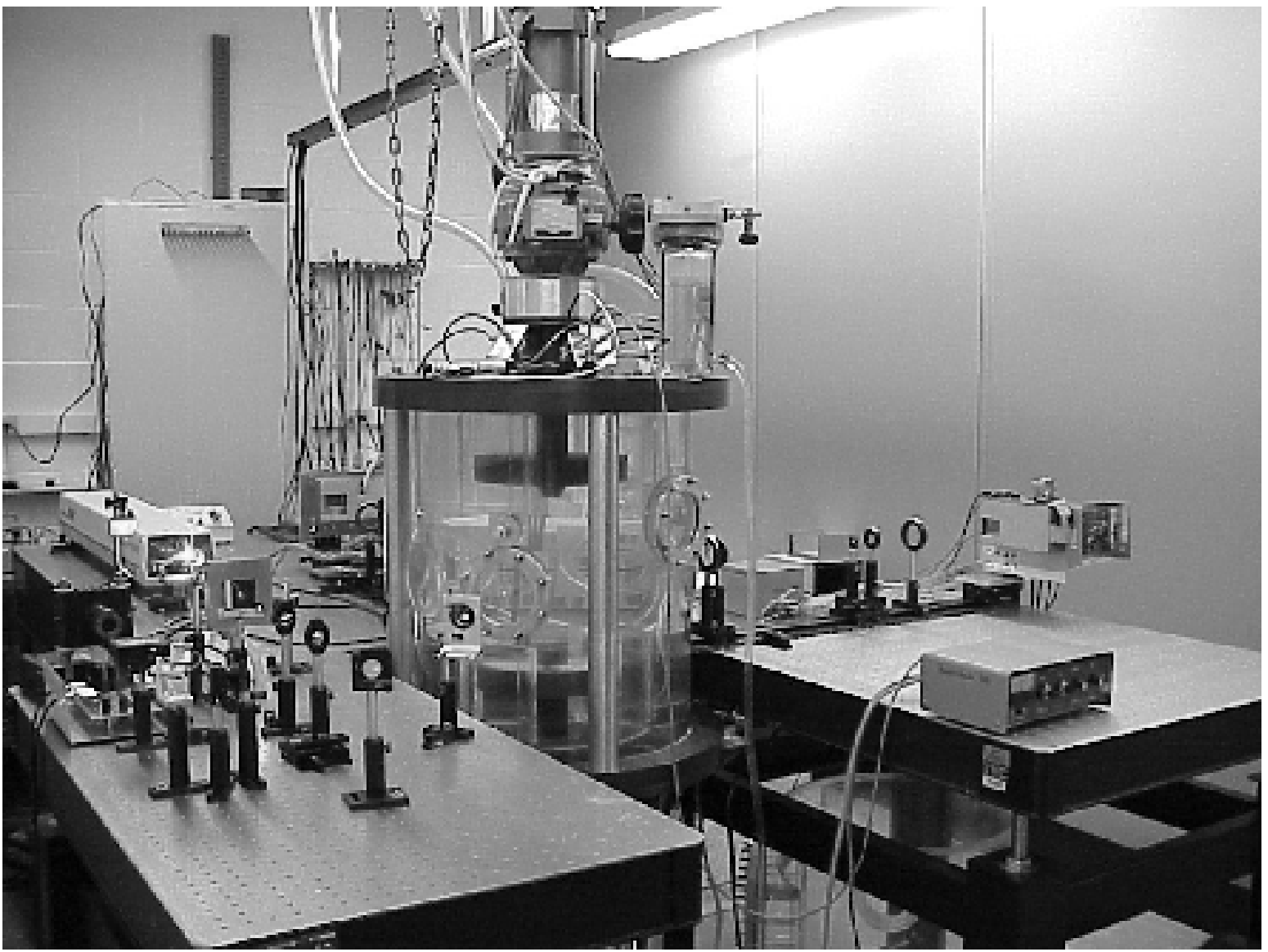}%
}
\end{center}
\caption{Apparatus, consisting of turbulence chamber with
counter-rotating disks (foreground), illumination beam entering from
foreground and two detectors (background).}
\label{fig:photo}
\end{figure}

The turbulence is generated in a flow of water between counter-rotating disks
in a cylindrical container \citep{Voth:1998:LAM,laporta:1999},
as represented schematically in figure~\ref{fig:flow} and shown in
figure~\ref{fig:photo}.
The container, mounted vertically, is 48.3~cm in diameter and 60.5~cm
long with 8 planar windows mounted flush to the surface of the cylinder
at equal angles along the mid-line of the chamber.
All physical quantities are defined with reference to a Cartesian coordinate
system in which the $z$ axis corresponds to the axis of symmetry of the
cylinder and the $x$
and $y$ axes correspond to the optical axes of the two imaging systems
represented in figure~\ref{fig:app}(b).
By symmetry the two transverse coordinates ($x$ and $y$) are equivalent,
and distinct from the axial coordinate ($z$).
The disk shaped propellers are open-ended cylinders 20~cm in diameter
and 4.3~cm deep with twelve internally mounted radial vanes.
The propellers are spaced 33~cm apart and are each driven by 0.9~kW computer controlled
controlled dc motors that are coupled to the propellers with variable
speed reducers.
A smaller cylindrical tube surrounds each propeller and stationary
radial vanes between this inner cylinder and the container wall have
been installed to inhibit large scale rotation of the flow.
For the studies described below, the propeller rotation rate is varied
from 0.15~Hz to 7.0~Hz.
The lower limit is set by the motors and speed reducers
and the higher limit is set
by the temporal resolution of the detector.
(The detectors are not able to adequately resolve the Kolmogorov
time at the maximum propeller speed of 9~Hz.)

The averaged flow produced by the propellers can be interpreted
as a superposition of two basic components, a pumping mode and a shearing
mode.
Centrifugal pumping by the propellers produces the flow represented
schematically in figure~\ref{fig:flow}(a).
The resulting mean strain field at the centre of the chamber (represented
by the arrows in figure~\ref{fig:flow}(a))
tends to enhance the axial component of the vorticity.
In addition, fluid near the top and bottom of the cylinder tends
to rotate collectively with the counter-rotating propellers,
creating a shear layer around the
edge of the flow midway between the propellers,
as represented in figure~\ref{fig:flow}(b).
Particle accelerations are measured in a 4~$\mathrm{mm}^3$ volume at
the centre of the flow chamber.
Quantitative measures of the flow, described in detail below,
are derived from measurements in this volume, and additional measurements
in a larger 15~mm~$\times$~30~mm volume.
More details regarding this flow are given in references.~\citep{Voth:1998:LAM}
and \citep{laporta:1999}. (See also Appendix~\ref{sec:old_data}.)

\subsection{Measurement of the Velocity Fluctuations}
\label{sec:velocity}

\begin{figure}
\begin{center}
\epsfxsize=\smallfigurewidth
\mbox{%
\epsffile{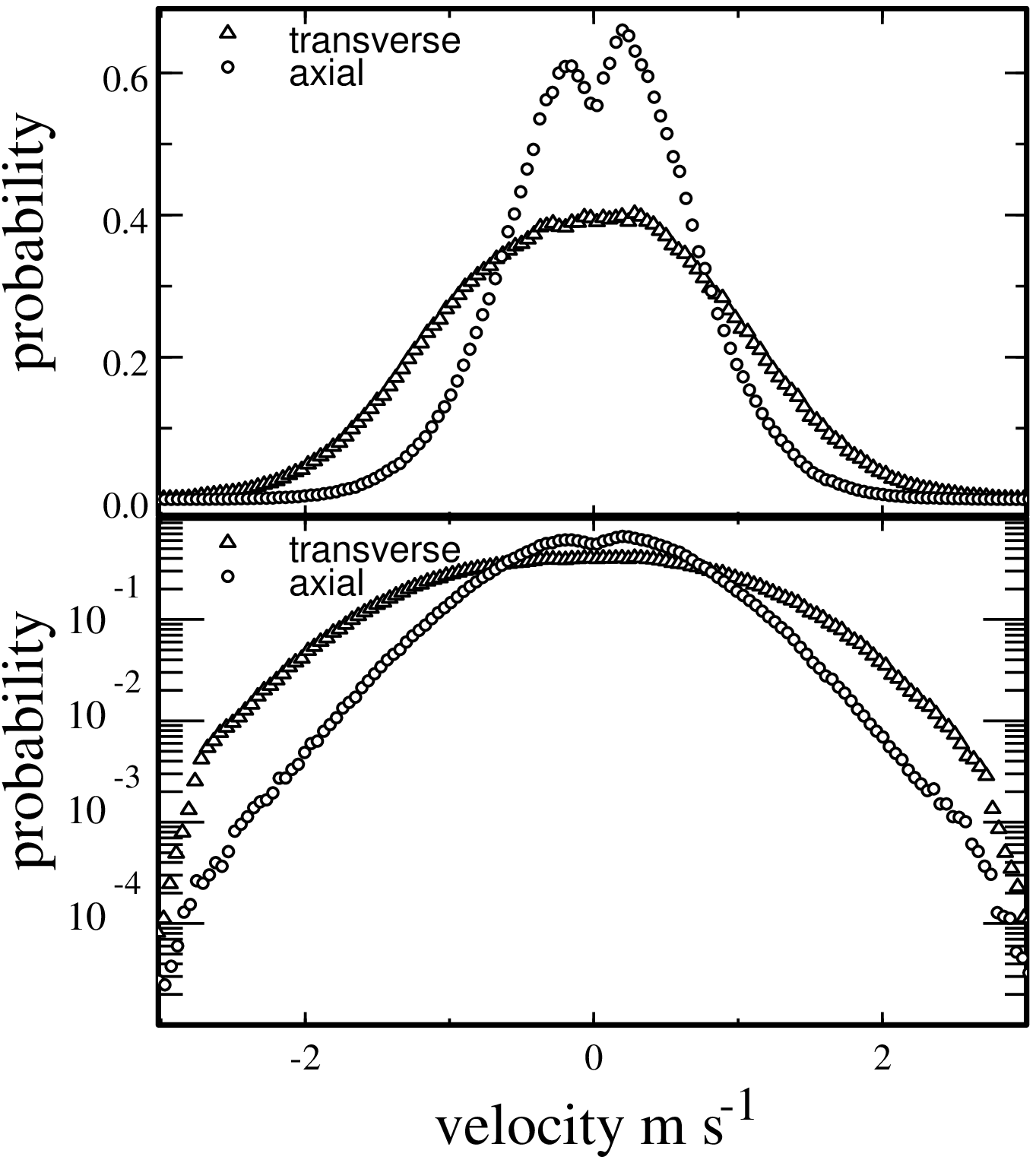}%
}
\end{center}
\caption{(a) The PDF for the velocity.
(b) Same data as (a) plotted on a semi-log scale.}
\label{fig:udist}
\end{figure}

Velocity statistics may be obtained from analysis of
matched trajectories obtained with the apparatus in its standard
configuration, shown in figure~\ref{fig:app}(a).
As mentioned above, in order to obtain a correct estimate of the
rms velocity, it is necessary to take care to sample trajectories
uniformly by insuring they are continuously sampled
as long as they remain in the measurement volume.
The PDF  of the velocity
is shown for linear and log scales in figure~\ref{fig:udist}.
It is evident that the standard deviation of the transverse component exceeds
that of the axial component by about 50\%.
The distributions for both components are approximately Gaussian; the
flatness is 2.8 for the axial and 3.2 for the longitudinal
component.  These flatness values are independent of the
propeller rotation frequency \citep{Noullez:1997:TVI}.
It may be noted that there is a small dip near zero velocity in the
axial velocity component PDF.  This occurs because the light sensitivity
of the detector decreases when a peak remains on the same pixel for
consecutive frames.  (This is due to an inefficiency in the shaping
amplifiers used for this detector.)  As a result, the measurement
volume is effectively smaller for trajectories with near-zero velocity,
causing a measurement bias.
It should be noted that measurement biases depend mostly on the velocity
and do not affect acceleration measurements, except to the extent that
the acceleration and velocity are correlated.

\begin{figure}
\begin{center}
\epsfxsize= \sfw
\mbox{%
\epsffile{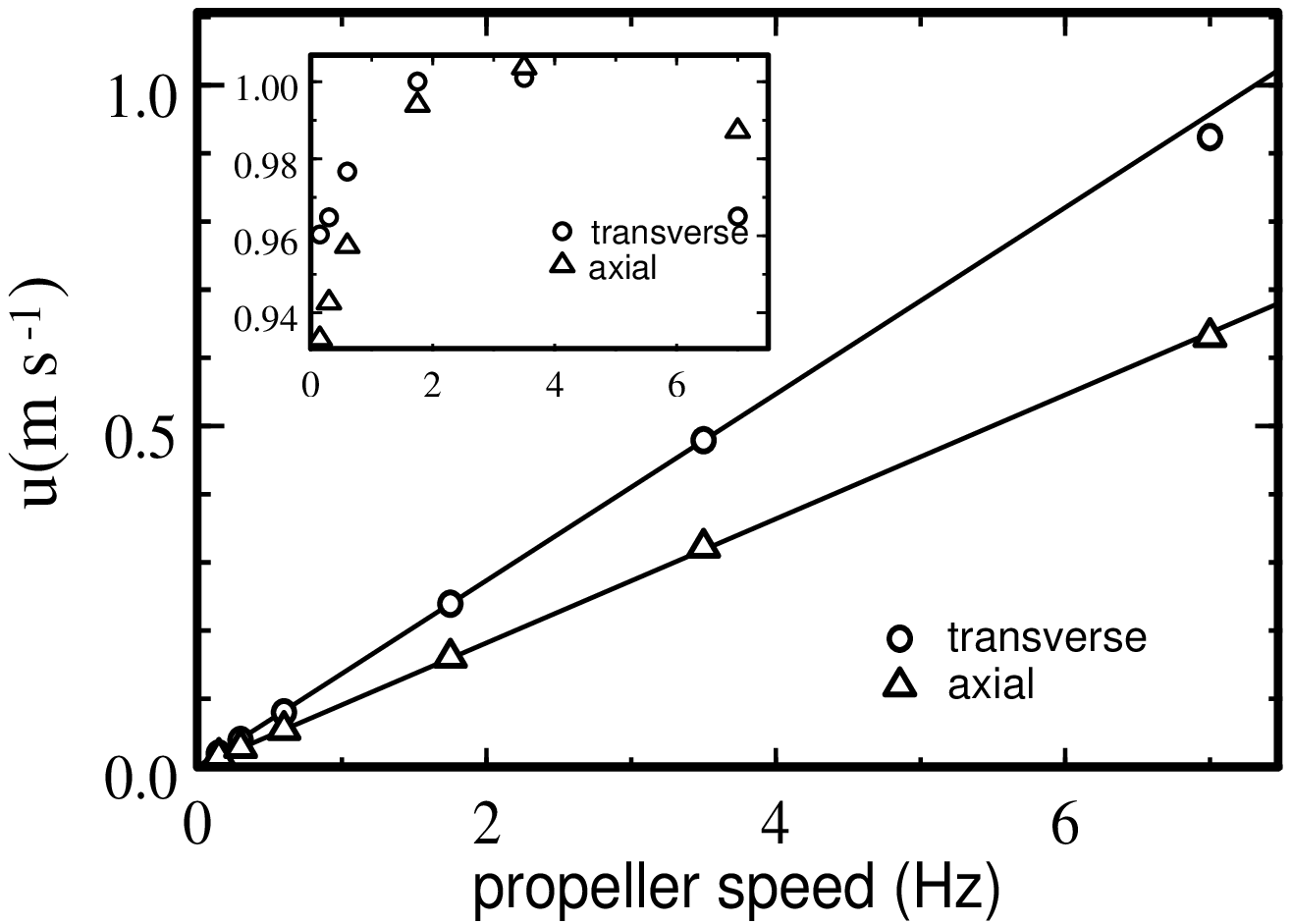}%
}
\end{center}
\caption{The axial and transverse components
of the rms velocity as a function of propeller speed.
The lines indicate the best fit to a linear relationship $u_i = k_i f$,
with $k_a = 0.0910$ for the axial component and $k_t = 0.137$ for
the transverse component.
The inset shows the
rms velocity components normalized by the linear scaling law.}
\label{fig:u_vs_f}
\end{figure}

The scaling of the rms velocity with propeller rotation frequency
is shown in figure~\ref{fig:u_vs_f}.
The expected linear dependence on the frequency
is observed for both components, which is consistent with the
assumption that the nature of the large scale flow is independent
of the stirring velocity over the range of Reynolds number studied.
The inset shows that the deviation from the linear scaling law
does not exceed a few percent.
This deviation is most pronounced at low propeller speeds, perhaps
indicating that the turbulence is not ``fully developed''
at the lower end of the Reynolds number range.

\begin{figure}
\begin{center}
\epsfxsize=\sfw
\mbox{%
\epsffile{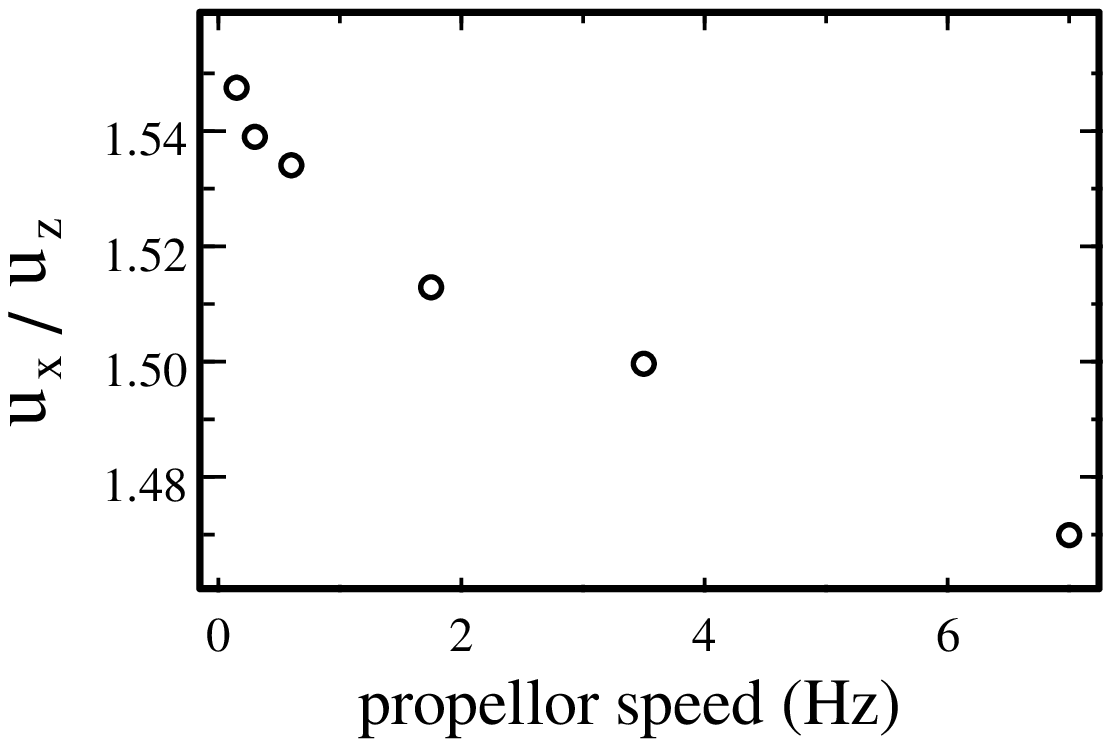}%
}
\end{center}
\caption{The ratio of transverse (x) to axial (z)  components of the rms velocity.}
\label{fig:uratio}
\end{figure}

The ratio of the rms transverse velocity to the rms axial velocity
is shown in figure~\ref{fig:uratio}.
The ratio varies only a few percent over the full range of
propeller speeds, again indicating that the large
scale structure of the flow does not change as the Reynolds number
is varied.

\subsection{Energy Dissipation}
\label{sec:dissipation}

In order to compare results with Kolmogorov scaling
predictions, it is essential to measure the energy dissipation
rate $\epsilon$, since all statistical quantities are assumed
to depend on this quantity and upon the kinematic viscosity $\nu$.
The energy dissipation $\epsilon$ is given by
\begin{equation}
\epsilon = 2 \nu \langle s_{ij} s_{ij} \rangle
\end{equation}
where $s_{ij}$ is the fluctuating rate of strain tensor, defined by
\begin{equation}
s_{ij} = \frac{1}{2} \left(
\frac{\partial u_i}{\partial x_j} +
\frac{\partial u_j}{\partial x_i}
\right),
\end{equation}
where $u_i = U_i - \langle U_i \rangle$ is a
component of the velocity fluctuation \citep{pope:TF}.
Unfortunately Lagrangian particle tracking does not allow us to
measure velocity gradients, so a direct measurement of the energy
dissipation is not possible.

There are several possible indirect methods for measuring the
energy dissipation from particle tracking data \citep{ott:2000}.
The most accessible method requires measurement of second (or third)
order velocity structure functions.
The velocity structure functions are the moments of the velocity
differences between two points separated by a fixed difference.
The longitudinal and transverse structure functions are calculated from
velocity components parallel to and perpendicular to the line separating
the two points, respectively.
From the second order structure functions, the energy dissipation
can be obtained by comparison with the K41 scaling relations,
\begin{eqnarray}
\dll&=&C_2 ( \epsilon r)^{2/3}\\
\dnn&=&\frac{4}{3} C_2 ( \epsilon r)^{2/3}
\label{eq:t2sf}
\end{eqnarray}
where $L$ and $N$ designate longitudinal and transverse, respectively,
and the separation $r$ is assumed to be within the inertial
subrange.
$C_2$ is an approximately universal constant that has been determined empirically
~\citep{Monin:1975:SFM,pope:TF}.

All of these structure functions can be
calculated from \tttd Lagrangian trajectory data by simultaneously measuring
the velocities of pairs of particles.
However, the particle tracking system used in this experiment
is currently not capable of \tttd tracking in a volume which
would encompass inertial
range particle separations due to power limitations of the
Argon-Ion laser used for illumination.
To overcome this, we have developed a technique for measuring the
transverse second order velocity structure function using \ttd
particle tracking in conjunction with a light sheet.

\begin{figure}
\begin{center}
\epsfxsize= \fw
\mbox{%
\epsffile{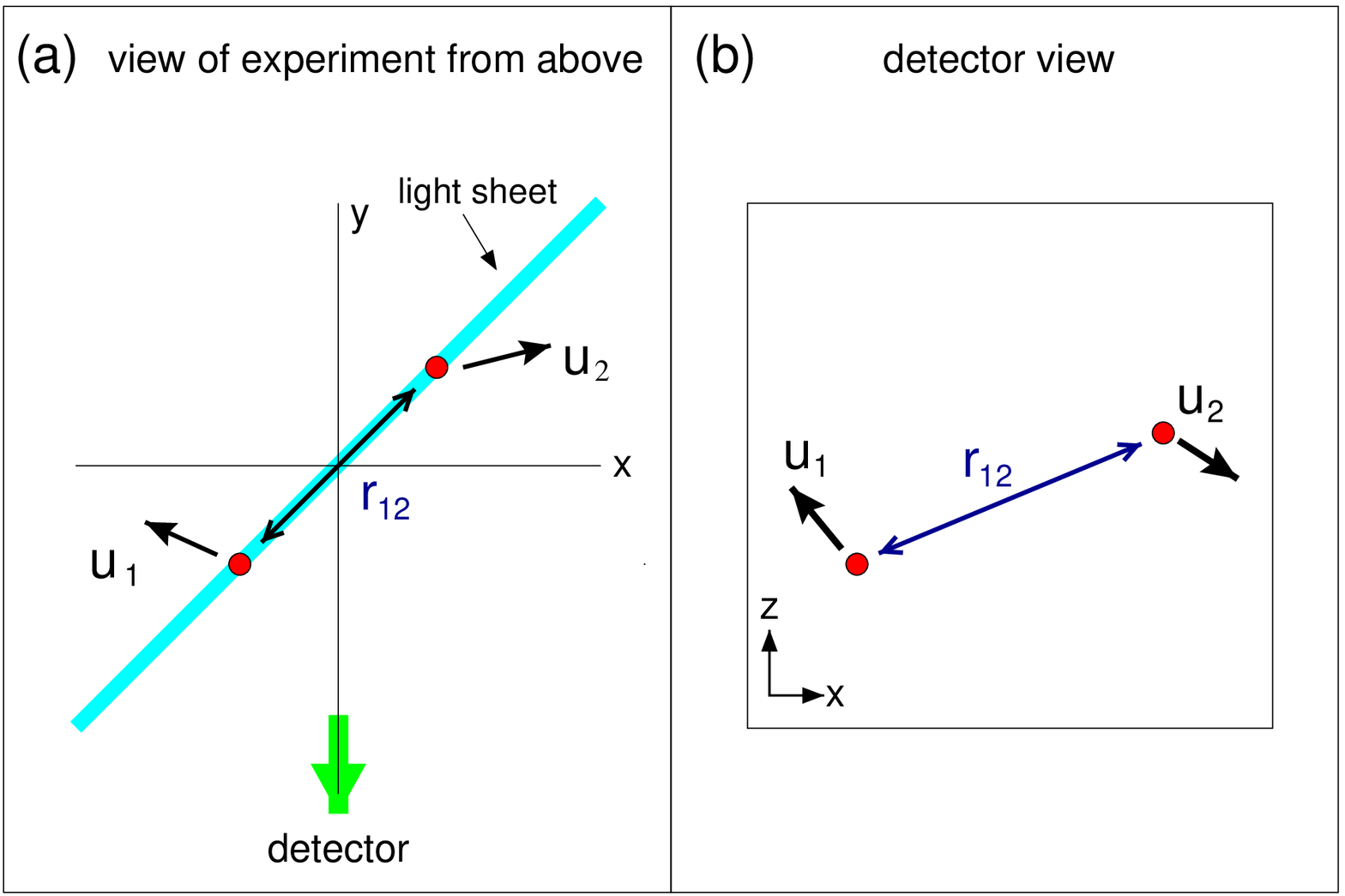}%
}
\end{center}
\caption{(a) Experimental setup for dissipation measurements viewed
from above (along the axial direction) showing the orientation of the
$45^\circ$
light sheet which illuminates the detection volume.
(b)Illustration of detector view for dissipation measurements.
The detector sees the particle separation and velocity projected on the $x-z$ plane, where
$z$ is the axial coordinate.
}
\label{fig:dis_config}
\end{figure}

The structure function measurements are performed using the
configuration shown in
figure~\ref{fig:dis_config}.
There are several significant differences between this
configuration and the standard
configuration used for single particle acceleration and velocity measurements.
The magnification has been reduced to 2.89, giving a field of view of
17.7~mm ($\ge 500\eta$), which allows inertial
range particle separations to be observed.
The optical system is schematically similar
to figure~\ref{fig:app}, except that the beam-splitter is omitted and
a single detector is used to measure tracks.
The bias of the detector is set so that conjugate peaks are enabled
and \ttd trajectories can be obtained from the single detector output,
as discussed in Section~\ref{sec:stripdetector} above.
In order to maintain adequate illumination intensity over such a large
field of view, the illumination beam is configured as a light sheet
approximately 0.1~mm thick and 10~mm wide.
The light sheet was created using a standard Galilean telescope to
expand the beam to a large $\mathrm{TM}_{00}$ mode,
then a second cylindrical telescope to compress the horizontal
axis and create an elliptical mode.
The measurement would be simpler in principle if the light sheet were
parallel to the image plane, but in practice it must be oriented
at $45^\circ$ with respect to the optical axis in order to
obtain sufficiently strong light scattering from the particles.
The data obtained from this configuration consists mostly of
short tracks which are created as the particles pass through
the light sheet.
The depth of the light sheet is chosen so that tracks have sufficient
length for an accurate velocity measurement to be made.

Velocity structure functions are measured as a function of
$r$ by calculating velocity
differences for all coincident pairs of particles
and compiling statistics under the condition that the
separation distance lies within an adjustable range.
Such a pair of particles will appear to the detector as shown in
figure~\ref{fig:dis_config}(b), and it is possible to measure
two velocity components for each particle and two components of the
particle separation vector.
However, as illustrated in figure~\ref{fig:dis_config}(a), the separation
vector does not lie in the same plane as the velocity components.
Using the coordinate system defined in
figure~\ref{fig:dis_config} the coordinates $x$ and $z$ are
measured explicitly and we may assume that $y \equiv x$.
The velocities are determined by tracking the particle
for the short time that it remains within the finite thickness
of the light sheet, and gives the projection of the velocity
on the imaging plane.  We therefore measure the components
$u_x$ and $u_z$, but $u_y$ is unknown.

\begin{figure}
\begin{center}
\epsfxsize= \bfw
\mbox{%
\epsffile{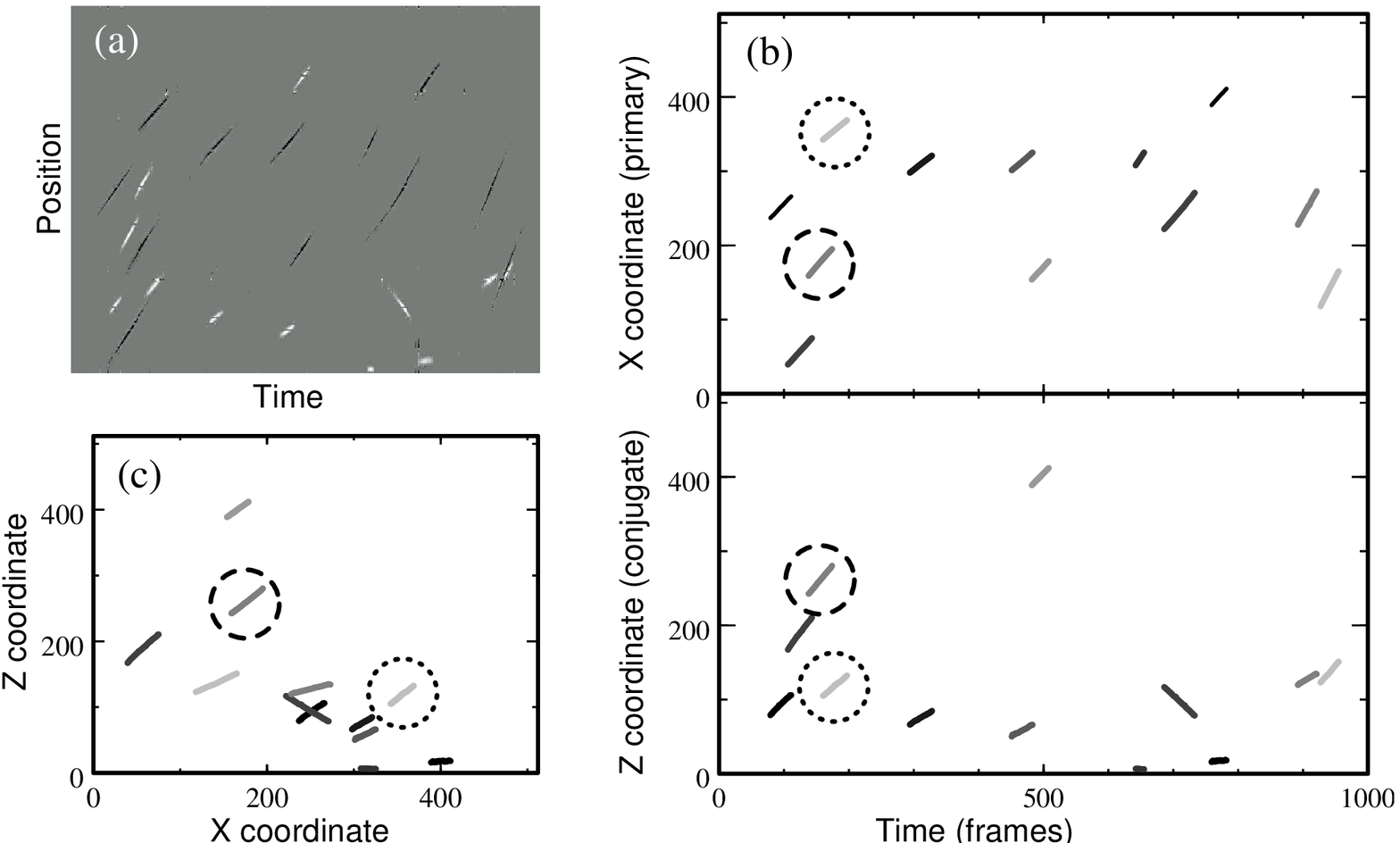}%
}
\end{center}
\caption{(a) Raw intensity data, in which dark areas indicate primary
(positive) charge and light areas indicate conjugate (negative) charge.
(b) Position vs time for $x$ (primary) and $z$ (conjugate) coordinates.
(c) Reconstruction of $x$-$z$ trajectories after track matching.
The dashed and dotted line circles indicate a pair of simultaneous tracks
from which velocity differences may be measured.}
\label{fig:dis_data}
\end{figure}

\begin{figure}
\begin{center}
\epsfxsize= \smallfigurewidth
\mbox{%
\epsffile{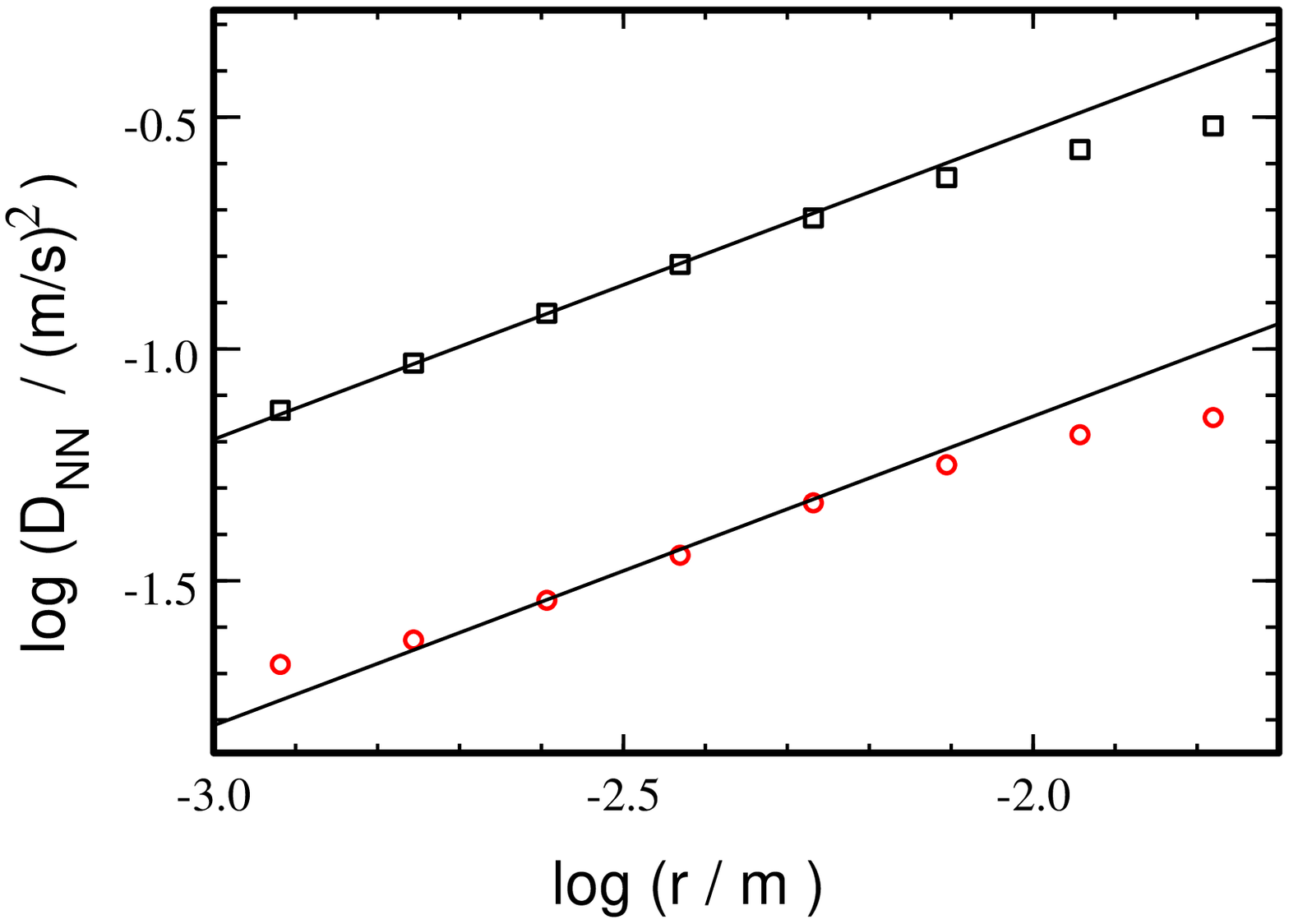}%
}
\end{center}
\caption{Log plot of the scaling of $\dnn$ with $r$.
The square symbols indicate data
at 5~Hz and the round symbols indicate data at 2.5~Hz. The straight lines
indicate best fit to $r^{2/3}$, as indicated in Table~\ref{tab:dis_param}.}
\label{fig:dis_scaling}
\end{figure}

Using this geometry, it is not possible to measure the longitudinal
structure function because it is in general not possible to measure
a velocity component along the separation vector.
However, it is always possible to measure
one of the velocity components perpendicular to the separation
vector.
To find the measurable transverse velocity component, we make use
of the fact that this component must be perpendicular to both the
separation vector $\br_{12}$ and to the vector
normal to the image plane ($\hat{y}$).
Such a vector may be constructed by taking the cross product of $\br_{12}$
and $y$.
The transverse velocity component is then
\begin{equation}
u_{\perp 2} - u_{\perp 1} = (\bu_2 - \bu_1) \cdot ( \br_{12} \times \hat{y} )/
\| \br_{12} \times \hat{y} \|,
\end{equation}
where $\bu_1$ and $\bu_2$ are the velocities of the two particles.
The dot product can be evaluated even though the $y$ component is unknown
because the $y$ component of $\br_{12} \times \hat{y}$ is identically
zero.

The structure functions are calculated from the second moment
of ($u_{\perp 2} - u_{\perp 1}$) conditional on $\br_{12}$,
\begin{equation}
\dnn(\bar{r}) =
\left\langle \left( u_{\perp 2} - u_{\perp 1} \right)^2 |
r < \| \br_{12}\| < (1+s) r
\right\rangle
\end{equation}
where the relative bin width $s$ is maintained constant as $r$
is varied and where $\bar{r}$ is the mean value of $\| \br_{12} \|$ for
all events which satisfy the condition.

Raw data used in the structure function are shown in
figure~\ref{fig:dis_data}(a) and $x$ and $z$ coordinate tracks
(calculated from primary and conjugate peaks, respectively) are shown
in figure~\ref{fig:dis_data}(b).
The matching of $x$ and $z$ tracks is performed using intensity
correlations (as described in Appendix~\ref{sec:trackextraction} below)
The results of the matching are shown in figure~\ref{fig:dis_data}(c).
The scaling of $\dnn$ with $r$ is shown in figure~\ref{fig:dis_scaling}.
Both exhibit the expected $r^{2/3}$ scaling over a substantial range.
The scaling range is limited at small $r$ because particles are too
close together to be well localized by the light sheet and at large
$r$ because their separation is approaching the integral length scale.
The energy dissipation rates obtained from the fits in
figure~\ref{fig:dis_scaling} are shown in Table~\ref{tab:dis_param}, and
and are consistent with the expected
$\epsilon = \tilde{u}^3/L$ scaling with $L = 7.1$~cm.
Once the energy dissipation has been determined, it is possible to
calculate the Taylor microscale,
$\lambda = ( 15 \nu \tilde{u}^2/\epsilon )^{1/2}$ and
the Taylor microscale Reynolds number
\begin{equation}
\Rl = \frac{\tilde{u} \lambda}{\nu}
= \left(\frac{15 \tilde{u}L}{\nu}\right)^{1/2}
= \frac{15^{1/2} \epsilon^{1/6}L^{2/3}}{\nu^{1/2}}
\label{eq:rlambda}
\end{equation}

\subsection{Mean Rate of Strain}
\label{sec:strain}

The optical configuration used for the dissipation measurements
is also useful for investigation of the structure of the large scale flow
near the centre of the apparatus.
Due to the symmetry at the centre of the flow, the off-diagonal
elements of the mean rate of strain tensor $\frac{\partial U_i}{\partial x_j}$
are zero, and the diagonal elements are constrained by symmetry in the
transverse plane and the incompressibility condition,
$\frac{\partial U_i}{\partial x_i} = 0$ (using the summation convention),
so that $\frac{\partial U_z}{\partial z} =
-2 \frac{\partial U_x}{\partial x} =
-2 \frac{\partial U_y}{\partial y}$.
Measurements at 5.0~Hz and 2.5~Hz show
a linear relationship
$\frac{\partial U_z}{\partial z} = \tilde{u}/0.0492\mathrm{m}$.
This may be compared with a component of the fluctuating strain,
which is related
to the dissipation in an isotropic flow by
$\epsilon = 15 \nu \left( \frac{\partial u_z}{\partial z} \right)^2$.
The ratio of the fluctuating strain to the mean strain is therefore
\begin{equation}
\frac{\left(\frac{\partial U_z}{\partial z}\right)}%
{\left(\frac{\partial u_z}{\partial z}\right)} =
\frac{(15 \nu L)^{1/2}}{0.0492\mathrm{m}}
u^{-1/2} = 0.021 u^{-1/2}.
\label{eq:Rl}
\end{equation}
This implies that at the maximum propeller speed of 7~Hz,
the mean strain is about 2\% of the fluctuating strain
and rises to about 15\% of the fluctuating strain at the minimum
propeller speed of 0.15~Hz.

\begin{table}
   \begin{center}
    \begin{tabular}{cccc}
         $f$&$\tilde{u}$&$\epsilon$&$L$\\
         (Hz)&($\mathrm{m} \, \mathrm{s}^{-1}$)&($\mathrm{m}^2 \, \mathrm{s}^{-3}$)&($\mathrm{m}$)\\\hline
          $2.5$&$0.3095$&$0.406$&0.073\\
          $5.0$&$0.6190$&$3.367$&0.070
      \end{tabular}
\end{center}
\caption{ Turbulence parameters for dissipation data. The integral
length scale $L$ is calculated using $\epsilon = \tilde{u}^3/L$
and $C_2 = 2.13$\protect \citep{sreeni:1995}. }
\label{tab:dis_param}
\end{table}

\begin{table}
\begin{center}
\begin{tabular}{ccccccccccc}
$f$&
$\tilde{u}$&
$\tau_s$&
$\epsilon$&
$\tau_\eta$&
$\eta$&
$\Rl$&
$\mathrm{Re}$&
$\Delta t$&
$\tau_\eta / \Delta t$&
$\tau_s / \tau_\eta$
\\
(Hz)&($\mathrm{m} \, \mathrm{s}^{-1}$)&($\mathrm{ms}$)&($\mathrm{m}^2 \, \mathrm{s}^{-3}$)&$(\mathrm{ms})$&
($\mu\mathrm{m}$)&&&($\mu \mathrm{s})$
\\\hline
$0.15$&$0.0186$&$107$&$9.01 \times 10^{-5}$&$105$&$322$&$140$&
$1,340$&$205$&$512$&$1.02$\\
$0.30$&$0.0371$&$53.9$&$7.21 \times 10^{-4}$&$37.0$&$191$&$200$&
$2,690$&$150$&$257$&$1.46$\\
$0.41$&$0.0509$&$39.7$&$1.85 \times 10^{-3}$&$23.1$&$151$&$235$&
$3,680$&$116$&$200$&$1.72$\\
$0.60$&$0.0743$&$26.9$&$5.77 \times 10^{-3}$&$13.1$&$114$&$285$&
$5,380$&$74.9$&$175$&$2.05$\\
$1.75$&$0.217$&$9.22$&$0.143$&$2.63$&$51.0$&$485$&
$15,700$&$25.9$&$102$&$3.50$\\
$3.5$&$0.433$&$4.62$&$1.14$&$0.929$&$30.3$&$690$&
$31,400$&$14.3$&$65$&$5.27$\\
$7.0$&$0.867$&$2.31$&$9.16$&$0.329$&$18.0$&$970$&
$62,700$&$14.3$&$23$&$6.99$\\
\end{tabular}
\label{tab:acc_param}
\caption{Turbulence parameters for acceleration data with
$\nu = 9.89 \times 10^{-7} \mathrm{m}^2 \,\mathrm{s}^{-1}$
(water at $20.6^\circ$).
$f$ is the propeller rotation frequency and
$\Delta t$ is the strip detector frame period.
$\tilde{u} = ((\bar{u}_x^2 + \bar{u}_y^2 + \bar{u}_z^2)/3)^{1/2}$
is the rms velocity.
The sweeping time $\tau_s$ is calculated from
$\tilde{u}$ and the 2.00~mm field of view of the detector.
The energy dissipation is
calculated from $\epsilon = u^3/L$ with $L = 0.071 \mathrm{m}$,
as discussed in Section~\ref{sec:dissipation}.
The Taylor microscale Reynolds number $\Rl$ is calculated
from equation~\ref{eq:rlambda} and the classical
Reynolds number $\mathrm{Re}$ is defined
by $\mathrm{Re} = \Rl^2/15$. }
\end{center}
\end{table}

\section{Results: Particle Acceleration Measurements}
\label{sec:results}

The primary subject of this paper is the study of fluid particle
accelerations in fully developed turbulence.
In contrast to the fluid particle velocity, which is the
same quantity that would be measured by a fixed probe at the same
location, the particle acceleration can only be measured using
Lagrangian techniques.
The acceleration of a fluid particle $\ba^+$ corresponds to the
substantive derivative of the velocity
\begin{equation}
\ba^+ \equiv
\frac{\partial \bu}{\partial t} + (\bu \cdot \nabla) \bu.
\end{equation}
In order to determine $\ba^+$ from Eulerian measurements, it would be necessary
to know $\partial \bu/\partial t$ as well $\bu$ and $\nabla \bu$
at a point in space (which is possible in DNS but not in experiments).
In terms of the fluid particle acceleration, the Navier--Stokes
equation is
\begin{equation}
\ba^+ = -\frac{\nabla P}{\rho} + \nu \nabla^2 \bu,
\end{equation}
where $P$ is the pressure and $\rho$ is the fluid density.
In fully developed turbulence, the viscous term is small compared
with the pressure gradient term, so a measurement of $\ba^+$
gives information about the pressure gradient, which is difficult
to measure experimentally.

Measurements of the acceleration require the lowest possible
position error, and were made exclusively from primary charge
readout (rather than from less accurate conjugate charge readout).
Accelerations were measured from \ttd trajectories recorded
in the configuration illustrated in figure~\ref{fig:app}(a).
Accelerations are therefore known in the $x$-$z$ plane,
giving one axial and one transverse component.
(Due to symmetry, the statistical properties of the unmeasured $y$ coordinate
are expected to be identical to those of the $x$ coordinate.)

\begin{figure}
\begin{center}
\epsfxsize=\fw
\mbox{%
\epsffile{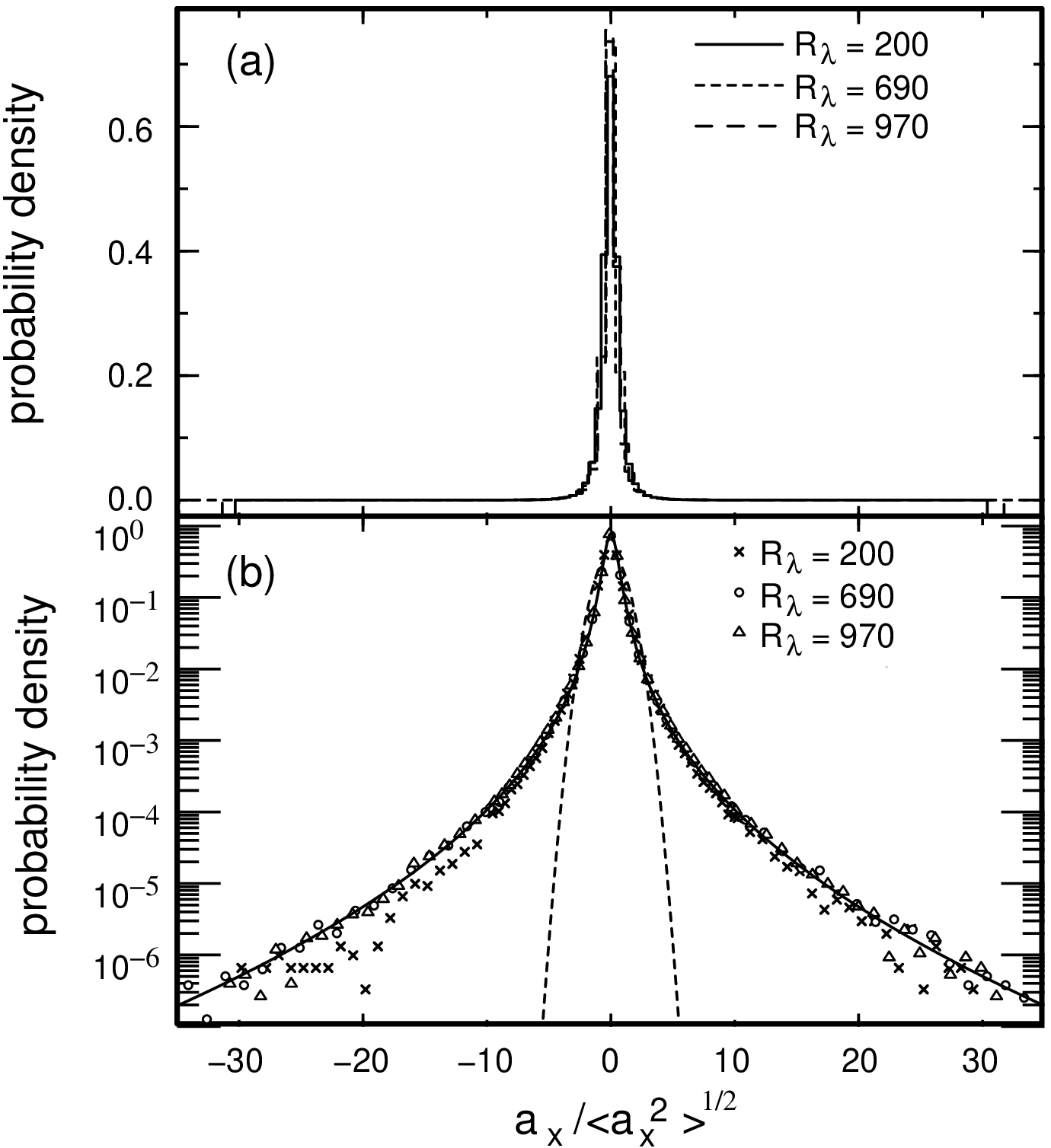}%
}
\end{center}
\caption{The PDF  of the transverse ($x$) component
of the acceleration at three values of $\Rl$ plotted on  (a) linear
and  (b) log scale.  Each acceleration distribution is measured by
parabolic fits over $0.75 \tau_\eta$ and is
normalized by its standard deviation. 
The dashed curve
is a Gaussian with the same standard deviation and the solid curve is the
parameterization defined in equation~\ref{eq:parameterization}.
}
\label{fig:adist}
\end{figure}

\subsection{Acceleration Distribution}
\label{sec:apdf}

The PDF  of the $x$ (transverse) component
of the acceleration is shown on linear and logarithmic scales
for several values of $\Rl$ in figure~\ref{fig:adist}.
It is found that the distributions have a stretched exponential form
for all measured values of $\Rl$, but that the extension of the
tails increases with $\Rl$.
The distributions are plotted for fits to a finite time interval
$\tau_f$ of $0.75 \tau_\eta$.
Although there is no qualitative change in the distribution as the
fit time is varied over a range $0.5\tau_\eta < \tau_f < 2 \tau_\eta$,
the moments of the distribution tend to increase
as the $\tau_f$ is reduced.
Estimates of the variance and flatness of the acceleration distribution
($\langle a_i^2\rangle$ and
$\langle a_i^4\rangle/\langle a_i^2\rangle^2$, respectively)
must therefore be obtained by measuring these quantities as a function of
$\tau_f$ and extrapolating to zero, as will be described in
Section~\ref{sec:avariance} below.
However, on the basis of figure~\ref{fig:adist} it
is obvious that the tails extend far beyond those of a Gaussian
distribution of the same variance and
that the flatness of the acceleration is
very large ($>40$, compared with $3$ for a Gaussian).
This result is consistent with the large
pressure gradient flatness values measured at low
Reynolds numbers in DNS\citep{vedula:1998}.

The acceleration component PDF may be parameterized by
the phenomenological function
\begin{equation}
P(a) = C \exp
( - [a^2]/[(1 + \left|a \beta/\sigma\right|^\gamma) \sigma^2] ),
\label{eq:parameterization}
\end{equation}
where $\beta = 0.539$, $\gamma = 1.588$, $\sigma = 0.508$ and $C = 0.786$
was obtained for the data at $R_\lambda = 970$.
It has been shown by \cite{holzer:1993}   that if the tails of the velocity difference distribution are
exponential, then the acceleration PDF should exhibit scaling
$P(a) \propto \exp(-a^{1/2})$. This may be compared with the tails of
equation~\ref{eq:parameterization} which have the form
$\exp( -|a|^{0.41} )$.

\begin{figure}
\begin{center}
\epsfxsize=\fw
\mbox{%
\epsffile{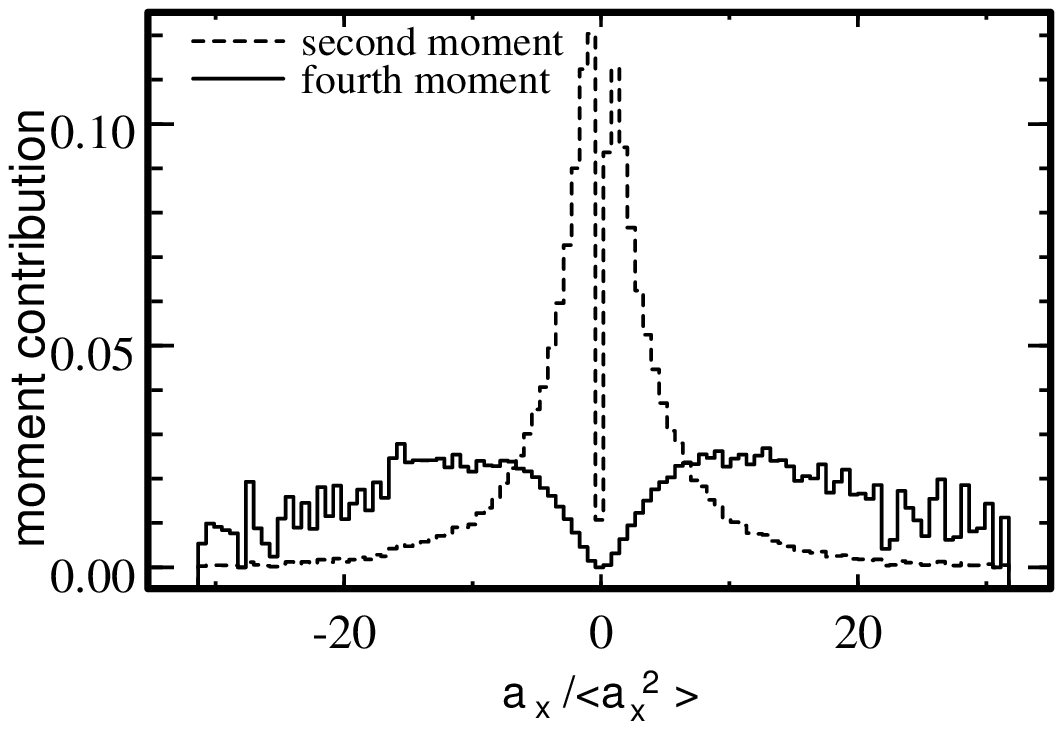}%
}
\end{center}
\caption{ The curves show the relative contribution to the
second and fourth moments
of the transverse acceleration component,
($a_x^2 P(a_x)/ \langle a_x^2 \rangle$
and ($a_x^4 P(a_x)/ \langle a_x^4 \rangle$) respectively,
as a function of acceleration. $P(a_i)$ is the PDF
of the $i$ component.
}
\label{fig:moment_contrib}
\end{figure}

As a result of the long tails of the acceleration PDF,
very large amounts of data are required
for convergence of the fourth moment, which is required for the calculation
of the flatness.  Figure~\ref{fig:moment_contrib} shows the contribution
to the second and fourth moments as a function of acceleration.  It
is apparent that even with $5\times 10^6$ acceleration measurements,
the convergence of the fourth moment is marginal.

\begin{figure}
\begin{center}
\epsfxsize= \sfw
\mbox{%
\epsffile{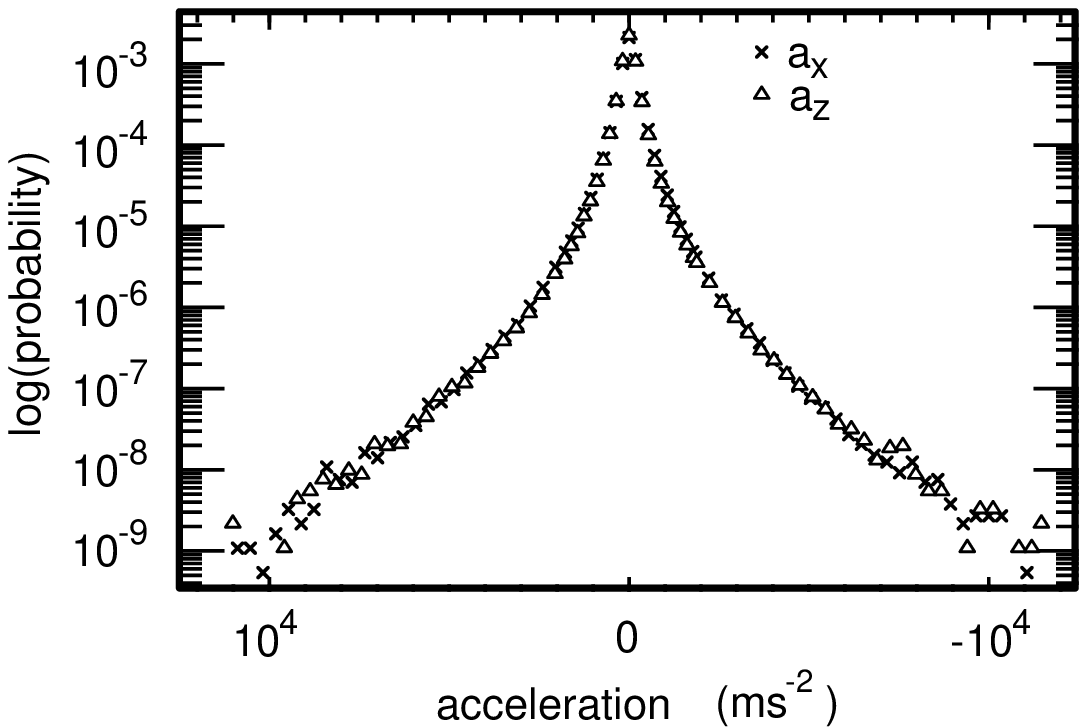}%
}
\end{center}
\caption{Acceleration distributions for transverse ($x$) and axial ($z$) components of
the particle acceleration for a run at $\Rl = 970$. The acceleration
was calculated for fits over 1~$\tau_\eta$.  The standard deviations
for $x$ and $z$ are 382~$\mathrm{m} \,\mathrm{s}^{-2}$ and
are 364~$\mathrm{m} \, \mathrm{s}^{-2}$ respectively.}
\label{fig:adist_7xy}
\end{figure}

The extent to which the anisotropy of the flow affects the acceleration
PDF is illustrated in figure~\ref{fig:adist_7xy}.
The plot shows the PDF's for $a_x$ and $a_z$ at $\Rl = 970$.
Comparison with figure~\ref{fig:udist} shows that the acceleration
is much less anisotropic than the velocity.
The small difference in the variances of $a_x$ and $a_z$ will
be discussed below in Section~\ref{sec:avariance}.

\subsection{Acceleration Variance}
\label{sec:avariance}

The variance of particle accelerations in a turbulent flow was first
predicted on the basis of the 1941 scaling theory of
Kolmogorov \citep{Kolmogorov:1941:LST,Kolmogorov:1941:DEL}
by Heisenberg and Yaglom \citep{heisenberg:1948,yaglom:1949}.
The  variance of the acceleration components is given by
\begin{equation}
\langle a_i a_j \rangle = a_0 \epsilon^{3/2}\nu^{-1/2} \delta_{ij}
\label{eq:ascale}
\end{equation}
where $a_0$ is predicted to be a  universal constant
 which is approximately 1 in a
model assuming Gaussian fluctuations \citep{heisenberg:1948}.
The form of this scaling law can be deduced from the assumption that
the acceleration is a dissipation scale quantity, and must be determined
only by $\epsilon$ and $\nu$.

Deviations from the Heisenberg-Yaglom scaling law are expected to
arise from turbulent intermittency. Using the refined similarity
theory, equation~\ref{eq:ascale} is replaced with
\begin{equation}
\langle a_i a_j \rangle = a_0 \nu^{-1/2} \delta_{ij}
\left\langle \epsilon_r^{3/2} \right\rangle,
\label{eq:rascale}
\end{equation}
where $\epsilon_r$, the energy dissipation averaged over a
sphere of radius $r$, has taken the place of the mean energy
dissipation, $\epsilon$.
Using the log-normal model for the moments of $\epsilon_r$,
this yields
\begin{equation}
\langle a_i a_j \rangle \propto (L/\eta)^{3 \mu_2/8}
\propto R_\lambda^{9\mu_2/16} = R_\lambda^{0.14}
\end{equation}
where a value of 1/4 has been used for the intermittency
exponent, $\mu_2$.
Other models of intermittency have been developed, such as
the explicitly Lagrangian model of Borgas \citep{Borgas:1993:MLN}, which
predicts $a_0 \propto \Rl^{0.135}$.

Direct numerical simulation of turbulence has shown that at low
Reynolds number $a_0 \propto \Rl^{1/2}$ (or $\epsilon^{1/12}$,
using equation~\ref{eq:Rl} and assuming constant $L$ and $\nu$).
This is equivalent to an overall scaling of the acceleration
variance with $\epsilon^{19/12}$, which is a relatively small
deviation from the Heisenberg-Yaglom prediction of
$\epsilon^{3/2}$.

\begin{figure}
\begin{center}
\epsfxsize=\smallfigurewidth
\mbox{%
\epsffile{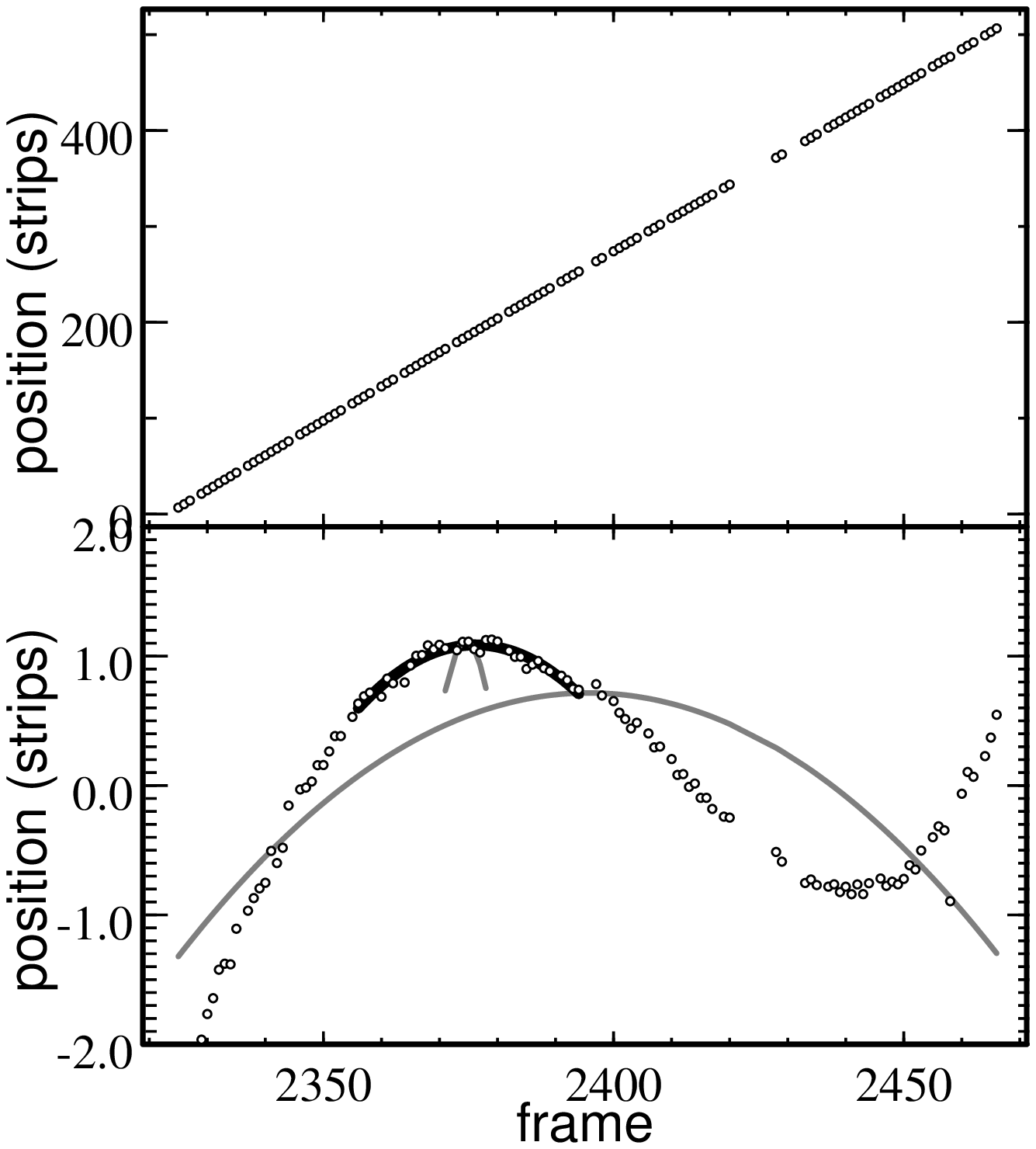}%
}
\end{center}
\caption{ Top panel shows an apparently straight particle trajectory
recorded at $\Rl = 970$.
The duration of the track is approximately 150 frames, 2~ms,
or 7~$\tau_\eta$.
The bottom panel shows the same trajectory after the mean velocity has
been subtracted off.
The three lines show parabolic fits over intervals of 150, 40
and 5 frames.}
\label{fig:data_fits}
\end{figure}

In principal, the acceleration variance may be calculated by taking the
second moment of the distributions shown in figure~\ref{fig:adist}.
However, in order to calculate the particle accelerations the track
must be fit over a finite time, and the moments of the distribution
depend on this fit time interval.
The issue is illustrated in figure~\ref{fig:data_fits}, which shows a
typical particle trajectory.
The raw trajectory appears to be straight, but when the mean velocity
is subtracted off, the particle is seen to be undergoing
a small time varying acceleration, which causes it to deviate from a straight
trajectory by a distance corresponding to a few strips on the detector.
The acceleration of this track is
approximately 90~$\mathrm{m} \, \mathrm{s}^{-2}$, which is about one quarter
of the rms acceleration---a typical value.
Clearly the correct acceleration will be measured from this
track only if a parabolic fit is made over an appropriate time interval.
The fit over 150 frames does not conform to the trajectory and underestimates
the acceleration.
The fit over 5 frames conforms to the position measurement errors
and dramatically overestimates the acceleration.  The fit over 40 frames
(1.8~$\tau_\eta$) appears to conform to this particular trajectory.

\begin{figure}
\begin{center}
\epsfxsize= \sfw
\mbox{%
\epsffile{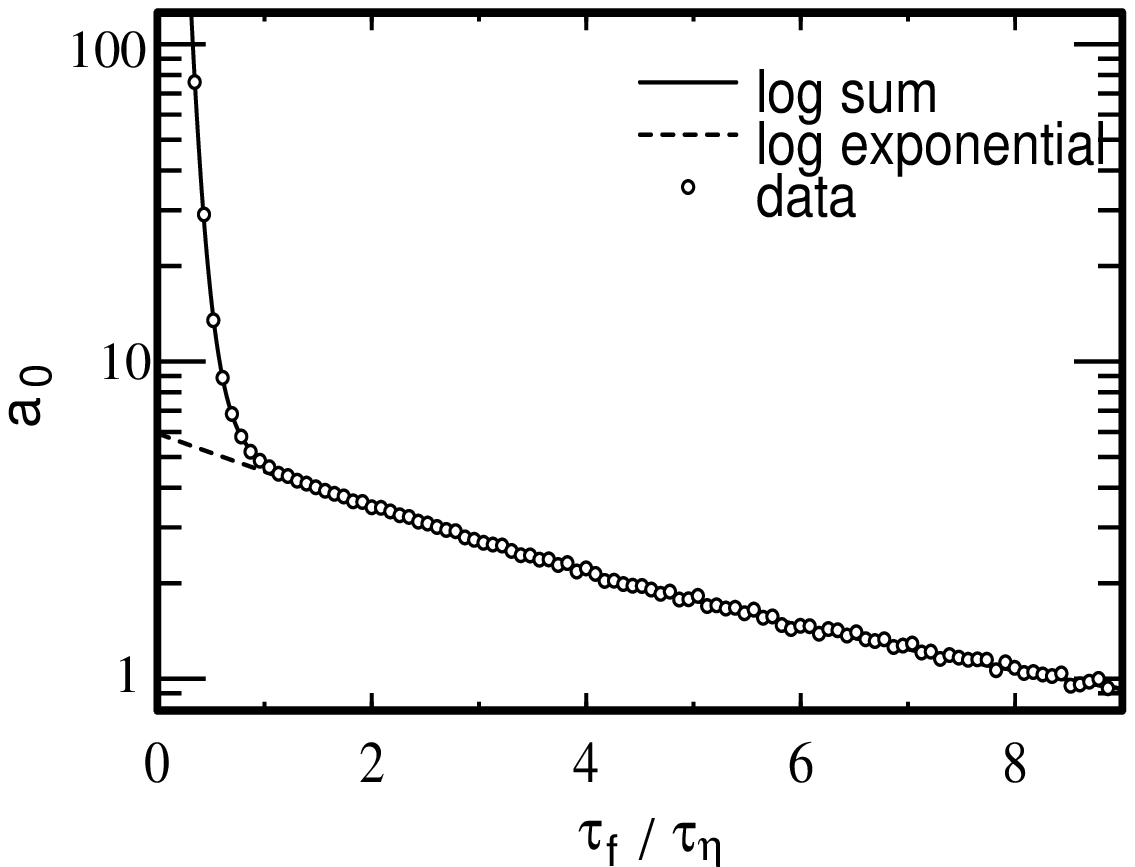}%
}
\end{center}
\caption{ Normalized acceleration variance ($a_0$) of x component as a function of fit
interval normalized by the Kolmogorov time ($\tau_f/\tau_\eta$).
The circles represent data taken at $\Rl = 970$, the solid
line indicates the best fit of the function in equation~\ref{eq:fit_fn},
and the dashed line shows the exponential term. A 10
extrapolation to zero fit time:  }
\label{fig:a0_fit}
\end{figure}

Ideally, we would like to fix the fit interval at a value where the
noise is adequately averaged and yet the parabolic fits are able to
conform to \textit{all} of the particle trajectories.
Figure~\ref{fig:a0_fit} shows the normalized variance of the acceleration
distribution as a function of the fit interval $\tau_f$,
and demonstrates that no value of $\tau_f$ exists which satisfies this
criteria,
since there is no range of $\tau_f$ where the
acceleration variance is independent of $\tau_f$.
For $\tau_\eta \le \tau_f \le 9\tau_\eta$ there is an
approximately
exponential dependence of the acceleration variance on the $\tau_f$
which is due to the failure of the fits to conform to the true
particle trajectories.
For $\tau_f < \tau_\eta$ the acceleration variance rises
dramatically with a $\approx \tau_f^{-5}$ power law dependence.
This is the $\tau_f$ dependence which would be obtained from uncorrelated
Gaussian distributed noise,
and evidently arises from the position measurement error.

The fact that the onset of position uncertainty occurs at a value of
$\tau_f$ where the fits fail to fully conform to the turbulent trajectories
indicates that
the frequency spectra of these two processes are not distinct, but overlap.
There is therefore no way to distinguish these two contributions on
any given track.
However, we can separate the two effects by making
use of the fact that the two contributions
to the acceleration variance have very different scaling with
$\tau_f$.
The procedure we use is to fit the measured $a_0$
(the normalized acceleration variance)
to the function
\begin{equation}
f(\tau) = A \tau^B + C \exp ( D \tau + E \tau^2 ),
\label{eq:fit_fn}
\end{equation}
where A, B, C, D and E are fit parameters.
The power law term represents the contribution from the
position noise and the exponential term represents the contribution
of the turbulence to the acceleration variance.
(The $\tau^2$ term is added to
model the slight deviation from exponential dependence observed at large
$\tau_f$.)
The best estimate of $a_0$ is then obtained by evaluating the exponential
term in the limit where $\tau_f \rightarrow 0$, so that $a_0 = C$.
It is known that this extrapolation must overestimate the value of $a_0$
because the slope of the $a_0$ vs $\tau_f$ curve must go to zero at
$\tau_f = 0$ since the tracks are differentiable.
A simulation of the detection process, described in
Section~\ref{sec:simulation} below, indicates that this overestimate
is 10\% at $\Rl \approx 240$, which is comparable to the random
measurement error.  All normalized acceleration variance data presented is rescaled to
correct for this overestimation.
This correction depends on the assumption that the dependence of
the measured value of $a_0$ on $(\tau_f / \tau_\eta)$ has a
universal form  for small $\tau_f$.

\begin{figure}
\begin{center}
\epsfxsize= \sfw
\mbox{%
\epsffile{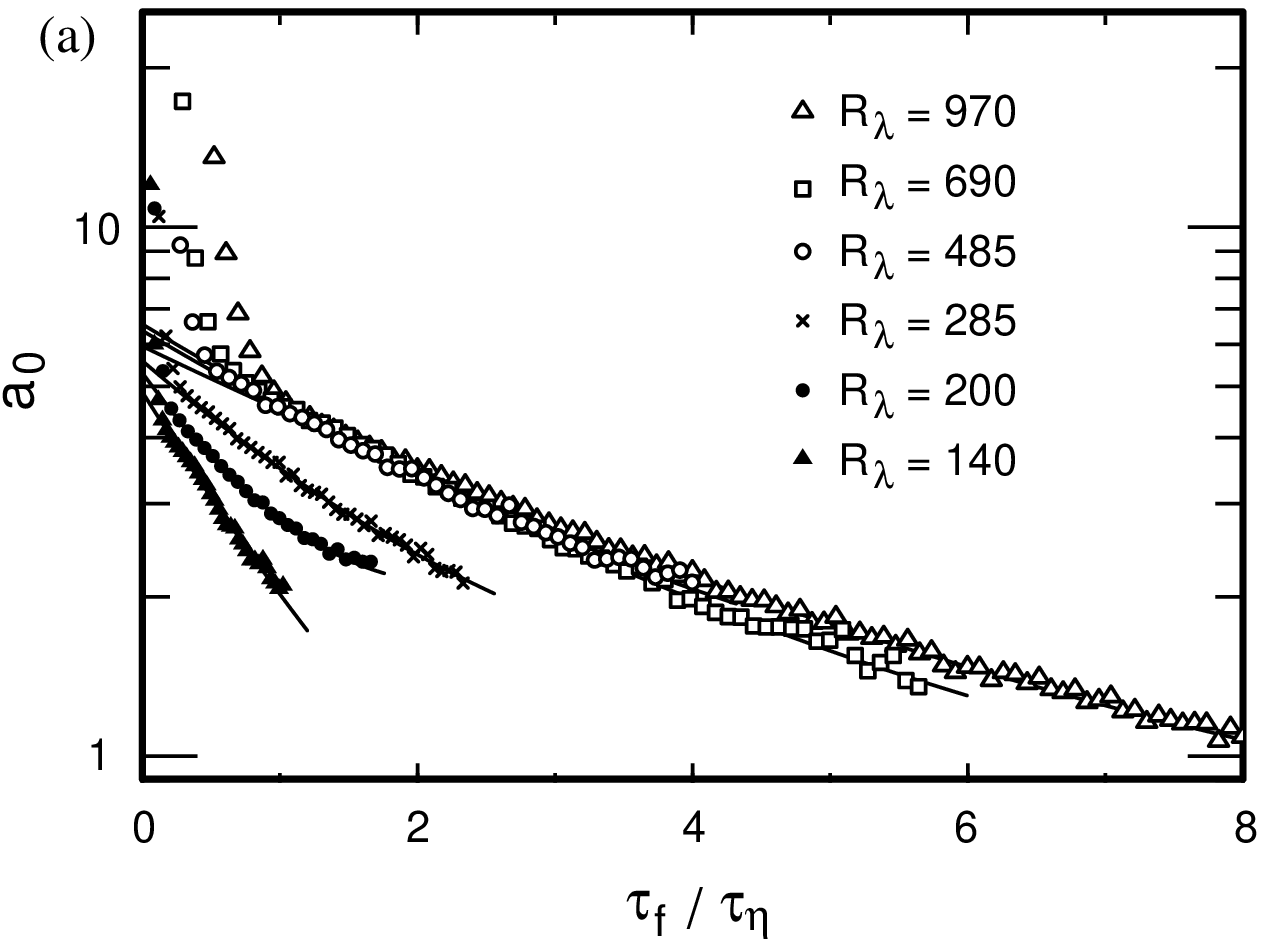}%
}
\epsfxsize= \sfw
\mbox{%
\epsffile{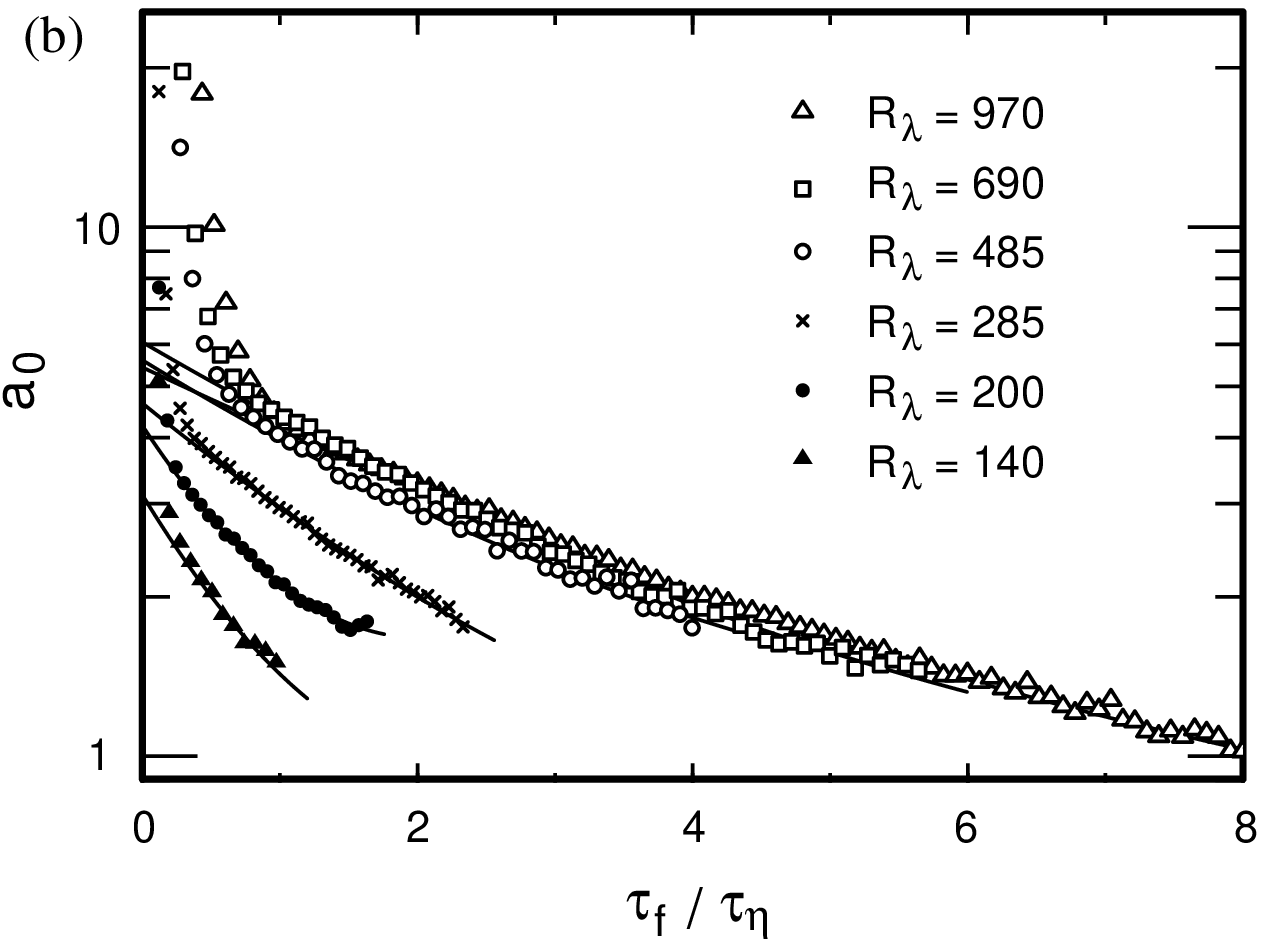}%
}
\end{center}
\caption{Plots of the normalized variance of an acceleration component,
$a_0$,
as a function of normalized fit time
$\tau_f/\tau_\eta$ for the full range of Reynolds number. The lines indicate
the exponential term $C \exp D \tau + E \tau^2$ from the best fit for
each curve.  The tabulated values of $a_0$ are derived from the extrapolation
of this term to $\tau = 0$. (a) x component (b) z component.}
\label{fig:a0x_vs_t}
\end{figure}

The procedure described above was used to calculate the acceleration
variance for $x$ and $z$ components of the acceleration for
all values of the Reynolds number.
Figures ~\ref{fig:a0x_vs_t} (a) and (b)
show $a_0$ as a function of $\tau_f$ for the $x$ and $z$ components
of the acceleration, respectively.
The maximum value of $\tau_f$ for which the acceleration variance can
be calculated is determined by the length of time the particles
remain in view.  This is estimated using the sweeping time $\tau_s$,
which is the length of time a particle moving at the rms velocity
would remain in the detection volume (tabulated in
Table~\ref{tab:acc_param}).
For purposes of fitting to equation~\ref{eq:fit_fn}
the range of each $a_0$ curve in
figure~\ref{fig:a0x_vs_t}
was limited at low $\tau_f$ to the value where the power law term
contributes an order of magnitude more than the exponential term,
and at high $\tau_f$ to $1.15 \tau_s$.

Certain conclusions may be drawn from figure~\ref{fig:a0x_vs_t}.
The curves for the three high Reynolds number runs collapse onto
a single curve, indicating that the scaling of the acceleration with
time is well approximated by K41 scaling.
At low Reynolds numbers, the
slopes of the curves and their extrapolations to $\tau_f = 0$
fail to collapse, indicating that K41 scaling has broken down.
However, the interpretation is somewhat more complex because
the measurement process brings an
additional time scale into play.
As the Reynolds number is reduced, the ratio of the
residence time of the particles $\tau_s$ to the
Kolmogorov time $\tau_\eta$ becomes smaller.
If there is any correlation between residence time and acceleration,
then this implies that
similarity can not be achieved by rescaling the time axis with $\tau_\eta$.
By varying the size of the measurement volume (and hence the sweeping
time $\tau_s$) we
have observed that the slope of the $a_0$ vs.\ $\tau_f$ curve
has some dependence on the finite size of the measurement
volume, although the extrapolation to $\tau_f = 0$ is found
to be independent of the measurement volume.
For this reason the $\tau_f \rightarrow 0$ extrapolation is a more
reliable measure of $a_0$ than its value at any finite $\tau_f$.

\begin{figure}
\begin{center}
\epsfxsize=\smallfigurewidth
\mbox{%
\epsffile{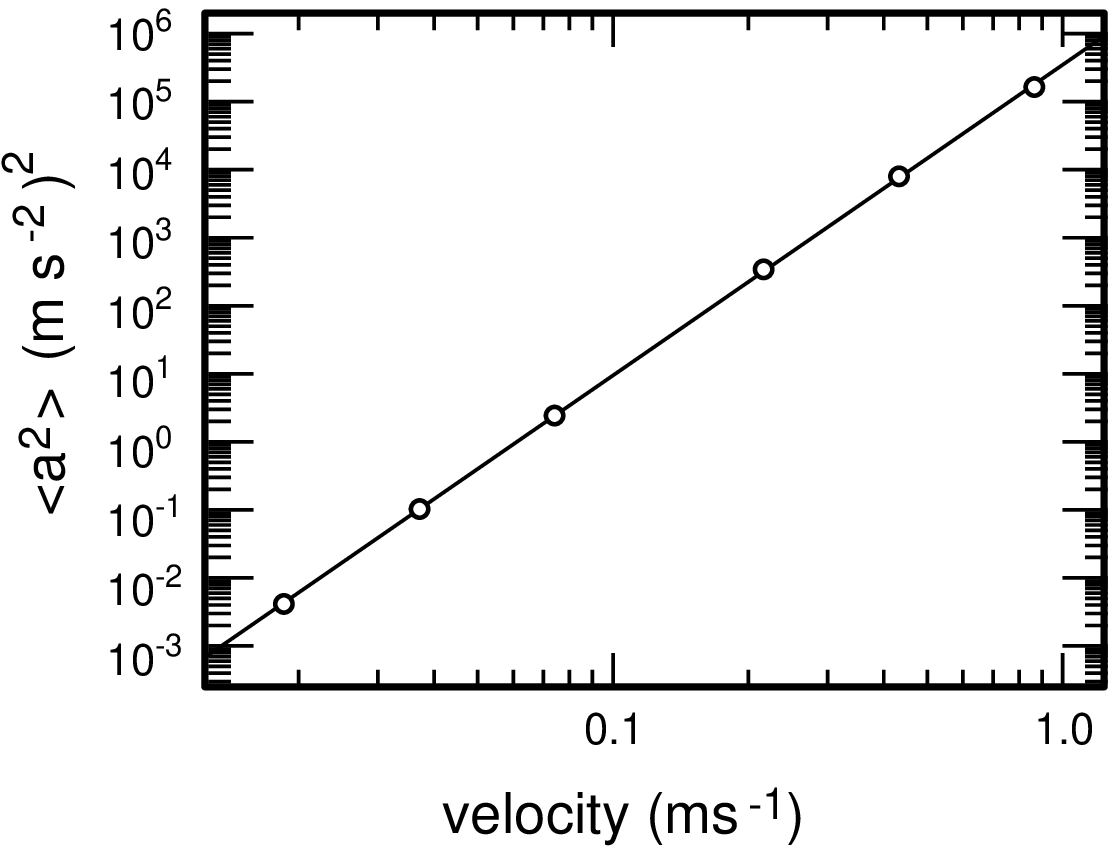}%
}
\end{center}
\caption{The variance of the acceleration plotted as a function of
$\tilde{u}$.  The straight line is the best fit of $\tilde{u}^{9/2}$
($\epsilon^{3/2}$ with $\epsilon = \tilde{u}^3/L$) to the data.}
\label{fig:a2_vs_u}
\end{figure}

\begin{figure}
\begin{center}
\epsfxsize=\smallfigurewidth
\mbox{%
\epsffile{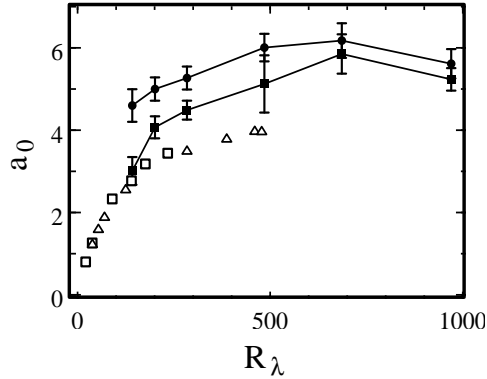}%
}
\end{center}
\caption{The Kolmogorov constant $a_0$ calculated for transverse (filled circles) and
axial (filled squares) components of the acceleration, as a function of Reynolds number.
These values have a 10\% correction applied to them as discussed in Section~\ref{sec:simulation}.
Values obtained from direct numerical simulation of turbulence by
Vedula and Yeung\protect \citep{vedula:1998},
and by Gotoh and Rogallo\protect \citep{gotoh:2001}
are also shown by open squares and open triangles, respectively.}
\label{fig:a0_vs_R}
\end{figure}

\begin{table}
\begin{center}
\begin{tabular}{ccccc}
$f $&$\Rl$&$(a_0)_x$&$(a_0)_z$&$(a_0)_x/(a_0)_z$\\
$(\mathrm{Hz})$& & & & \\ \hline
$0.15$&$140$&$4.60\pm0.40$&$3.01\pm0.27$&$1.53$\\
$0.30$&$200$&$4.99\pm0.28$&$4.07\pm0.48$&$1.23$\\
$0.60$&$285$&$5.26\pm0.28$&$4.49\pm0.70$&$1.17$\\
$1.75$&$485$&$6.01\pm0.33$&$5.12\pm0.23$&$1.17$\\
$3.50$&$690$&$6.18\pm0.42$&$5.85\pm0.27$&$1.06$\\
$7.00$&$970$&$5.61\pm0.35$&$5.23\pm0.34$&$1.07$
\end{tabular}
\label{tab:a0}
\caption{$a_0$ as a function of Reynolds number
for transverse and axial acceleration.}
\end{center}
\end{table}

The variance of the (unnormalized) $x$ component of the acceleration
is shown as a function of the rms velocity $\tilde{u}$ in
figure~\ref{fig:a2_vs_u}.
It is found that the predicted scaling
\begin{equation}
\langle a_x^2 \rangle = \frac{a_0 \tilde{u}^{9/2}}{L^{3/2}\nu^{1/2}}
\end{equation}
(From equation~\ref{eq:ascale}, using $\epsilon = \tilde{u}^3/L$),
is observed over nearly 7 orders of magnitude in acceleration
variance, or nearly 2 orders of magnitude in velocity variance. The scaling
of $a_0$ with $R_\lambda$ is plotted in figure~\ref{fig:a0_vs_R}
and (tabulated in Table~\ref{tab:a0}). In this plot, a constant
value of $a_0$ would indicate Heisenberg-Yaglom scaling. As $\Rl$
is decreased below 500 the value of $a_0$ decreases substantially
showing a small but significant departure from universal scaling.
The dependence of $a_0$ on $\Rl$ in this regime seems to be
qualitatively consistent with DNS results. It is not surprising
that DNS and experimental results do not match exactly at small
Reynolds number, since in this range the acceleration is coupled to 
 the large scales of the flow. The DNS results were
obtained for isotropic turbulence with periodic boundary
conditions, which differs markedly from the anisotropically forced
turbulence between counter-rotating disks used in the experiment.
At high Reynolds number ($\Rl \ge 500$) $a_0$ appears to be
independent of $\Rl$, which is consistent with Heisenberg-Yaglom
scaling. Due to experimental uncertainties, very weak deviations such
as the $\Rl^{0.135}$ prediction of the Borgas multi-fractal model
cannot be ruled out by this data.

Please note that it has been shown that the mean squared pressure gradient in a
turbulent flow---and
therefore the mean squared acceleration, and $a_0$---
is closely related to the fourth order velocity
structure functions and to the
inertial range flatness factor \citep{hill:1995}.
This would indicate that the
$a_0$ vs $\Rl$ curve in figure~\ref{fig:a0_vs_R}
could be compared with the flatness factor ($F$) vs $\Rl$ curve
reported by Belin et.\ al.\ in another experiment involving
a flow between counter-rotating disks \citep{belin:1997}.
Some similarity is evident, despite the fact that the flow configurations
are quite different ~\citep{hill:private}.

\begin{figure}
\begin{center}
\epsfxsize=\smallfigurewidth
\mbox{%
\epsffile{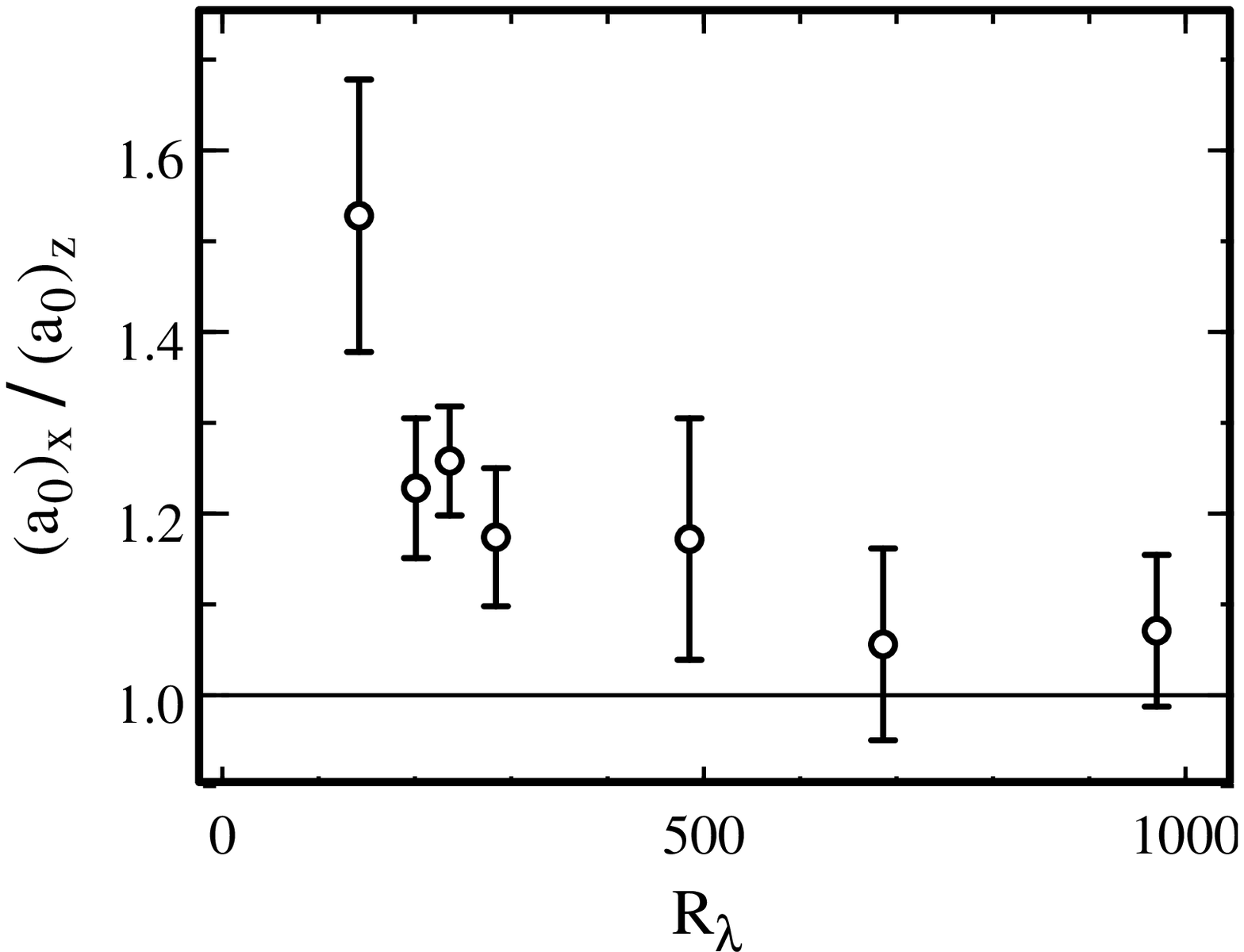}%
}
\end{center}
\caption{The ratio of $a_0$ calculated for transverse ($x$) and
axial ($z$) components of the acceleration as a function of Reynolds
number.}
\label{fig:a0_ratio}
\end{figure}

It is also evident that larger $a_0$ values are obtained for
the transverse component than for the axial component of the acceleration,
and that the level of anisotropy
decreases as the Reynolds number is increased.
The anisotropy of the acceleration variance is best illustrated by
calculating the ratio $(a_0)_x/(a_0)_z$ as a function of
$\Rl$, as shown in figure~\ref{fig:a0_ratio}.
At the lowest Reynolds number of 140,
$(a_0)_x/(a_0)_y = 1.52$, indicating that the ratio
of the standard deviations of the acceleration components is 1.23.
As the Reynolds number is increased to 970, the level of anisotropy
decreases to a very small value.
This observation is consistent with recent experimental results
which indicate that anisotropy persists to high Reynolds
numbers \citep{sreeni:2000,warhaft:2000}

It is also informative to compare figure~\ref{fig:a0_ratio} with
the inset to figure~\ref{fig:u_vs_f}, which shows the ratio of the
standard deviations of the velocity components.
On the one hand, the particle velocities are associated mainly with the
large scales and the level of anisotropy is large and independent of
Reynolds number, as expected.
On the other hand, the particle accelerations come mainly from the dissipation range
scales. As expected, the level of anisotropy for the acceleration is smaller than
for the velocity, and decreases as the Reynolds number is increased.

\subsection{Acceleration Flatness}
\label{sec:aflatness}

It is also interesting to quantify the degree of intermittency
of the acceleration by calculating the acceleration flatness
($\langle a_i^4 \rangle/ \langle a_i^2 \rangle^2$)
as a function of Reynolds number.
In this case, one encounters the same difficulty as for the variance;
the flatness varies as a function of the interval over which the
acceleration is calculated.
One must also confront the added difficulty that the tails of the
acceleration distribution are so long that prohibitively large
data samples would be needed to definitively converge the fourth
moment, as illustrated in figure~\ref{fig:moment_contrib} above.
In view of figure~\ref{fig:moment_contrib}, it seems possible
that the unconverged tails of the
acceleration distribution could make substantial contributions to the
fourth moment.
In addition, the ability of the tracer particles to fully follow rare violent events which
contribute most to the flatness has not been fully established.
Therefore, we can only set lower bounds to the flatness.

\begin{figure}
\begin{center}
\epsfxsize= \sfw
\mbox{%
\epsffile{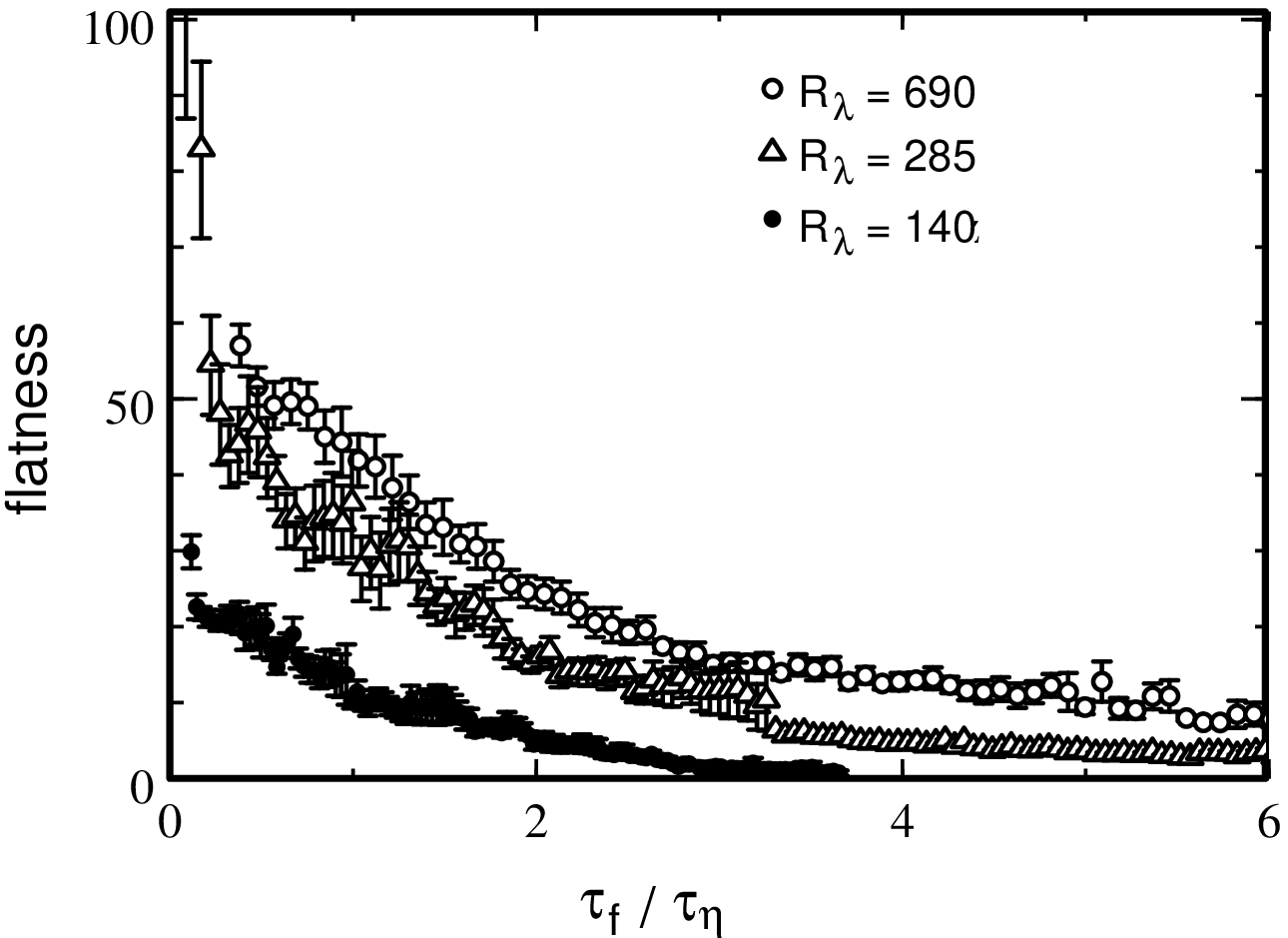}%
}
\end{center}
\caption{Flatness of the transverse component of the acceleration as a
function of fit interval $\tau_f$.}
\label{fig:xflatness}
\end{figure}

\begin{figure}
\begin{center}
\epsfxsize= \sfw
\mbox{%
\epsffile{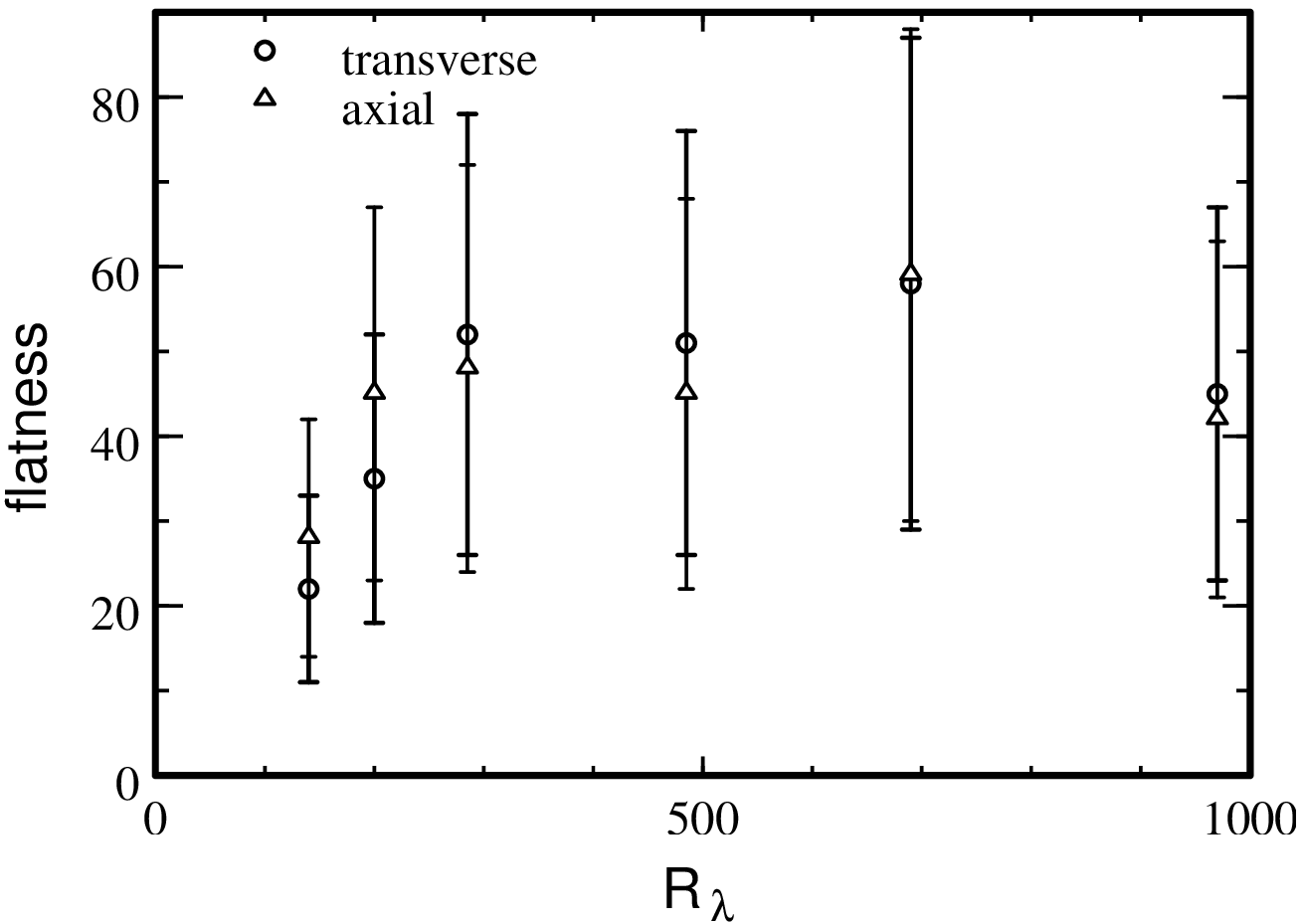}%
}
\end{center}
\caption{Lower bound on the flatness of the transverse
and axial components of the acceleration as a function of $\Rl$.
The error bars indicates an estimated uncertainty due to the
inability to extrapolate $\tau_f$ to zero,
but do not reflect uncertainty in the convergence of the
fourth moment.}
\label{fig:f_vs_r}
\end{figure}

The flatness for the transverse component of the acceleration
is shown as a function of fit time in figure~\ref{fig:xflatness}.
An increasing trend is evident as $\tau_f$ decreases, but the dependence
on $\tau_f$ is more difficult to define than in the analogous curves
for the acceleration variance
(figure~\ref{fig:a0x_vs_t}).
We have not found it possible to make a formal extrapolation for
$\tau_f \rightarrow 0$ in this case,
but by tabulating the flatness at a value of $\tau_f$ slightly
above the onset of power-law position uncertainty we can
compile a plot of lower bounds on the acceleration flatness, shown
in figure~\ref{fig:f_vs_r}.
The flatness is found to be at least 25 at the lowest Reynolds numbers
studied and the lower bound increases to approximately 60 as
the Reynolds number is increased.

These large flatness values indicate that the
acceleration is more intermittent than the other
small scale quantities in turbulence.
For example, at $R_\lambda = 200$, the longitudinal velocity
derivative flatness is 6.0 \citep{vanatta:1980}, and the scalar gradient flatness
is 17 \citep{warhaft:2000}, while the acceleration flatness in both experiments
and simulation \citep{vedula:1998} is approximately 30.

\subsection{Acceleration Correlations}
\label{sec:acorrelations}

Having measured the acceleration as a function of time
for the particle tracks, it is straightforward to evaluate
the acceleration autocorrelation function,
\begin{equation}
C_{a^+}(\tau) = \frac{\left\langle a_i^+(t) a_i^+(t+\tau) \right\rangle}
{
\left(\left\langle a_i^+(t)^2\right\rangle
\left\langle a_i^+(t+\tau)^2\right\rangle\right)^{1/2},
}
\end{equation}
where $a_i^+(t)$ is the acceleration component along a particle trajectory
and $\langle \rangle$ denotes averaging over $t$ for an ensemble of
tracks.
In principle, the denominator can be simplified and expressed as
the variance of $a_i^+$
but because the lengths of the tracks are of the same order as the
range of $\tau$, end effects can be significant;
$\left\langle a_i^+(t)^2\right\rangle$
and $\left\langle a_i^+(t+\tau)^2\right\rangle$ are compiled from
different data samples and can differ slightly,
particularly at large $\tau$.

\begin{figure}
\begin{center}
\epsfxsize=\smallfigurewidth
\mbox{%
\epsffile{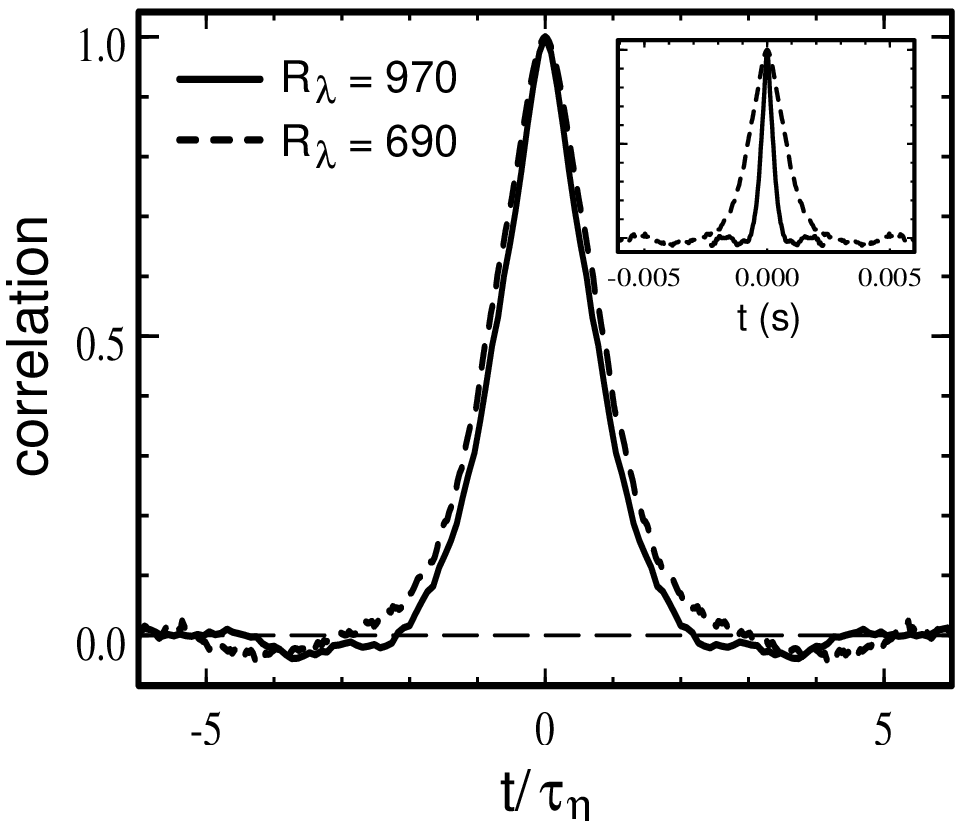}%
}
\end{center}
\caption{The acceleration autocorrelation function is shown for
data at $\Rl$ of 970 and 690, where the time axis has been
scaled with the Kolmogorov time $\tau_\eta$.
The accelerations are calculated by direct differencing of trajectories
which have been smoothed with a Gaussian kernel,
where the standard deviation of the kernel is $0.5~\tau_\eta$.
The insert shows the same data in which the time is unscaled
and plotted in seconds.}
\label{fig:acorr}
\end{figure}

The acceleration autocorrelation is shown for $\Rl = 970$ and
$\Rl = 690$ in figure~\ref{fig:acorr}.
In the main graph the time axis is normalized by the Kolmogorov
time $\tau_\eta$ for each curve, and in the inset the unnormalized time axis
is in units of seconds.
The Kolmogorov time is proportional to $\tilde{u}^{-3/2}$ and changes
by a factor of $2\sqrt{2}$ between the two values of $\Rl$
studied.
The data reproduces this scaling to an accuracy of a few percent.
Calculations of the acceleration autocorrelation function in
direct numerical simulation of turbulence at $R_\lambda = 140$ indicate that it
crosses zero at $2.2 \tau_\eta$ \citep{Yeung:97}.
In our experiment we see the zero crossing at $2.1 \tau_\eta$ at $\Rl = 970$
and $3 \tau_\eta$ at
 $\Rl = 690$.
The data collapse shown in figure~\ref{fig:acorr}
 may be regarded as an independent confirmation
of the measurement of the energy dissipation, described above
in Section~\ref{sec:dissipation}.

The autocorrelation data is compiled using the same trajectories as were
used for the study of the acceleration moments.
This presents some difficulty because of the small size of the measurement
volume.
The autocorrelation function is compiled under the implicit assumption
that the particle remains in view after a time interval $\tau$ has
elapsed, however,
particles with large accelerations are less likely to remain in view
than particles with small accelerations.
The ensemble of measurements contributing
to autocorrelation function at $\tau = 5 \tau_\eta$ have a standard
deviation that is a factor of 0.7
smaller than that of the unconditional acceleration variance, indicating
that the bias towards low acceleration events is appreciable.
To make a definitive study of the acceleration autocorrelation,
and particularly to obtain results at low Reynolds number, a more powerful
laser which can illuminate a larger measurement volume will be required.

\subsection{Simulation of Detection process.}
\label{sec:simulation}
As discussed before,
measurements of Lagrangian statistics are complicated by the
fact that all statistics are implicitly conditional on the
event being successfully measured.
For measurements of acceleration, the particle must remain
in view for a sufficient time for its acceleration to be
determined, which may introduce biases due to measurement
volume effects.
The extrapolation to zero fit time further complicates the
matter, because there are two effects which may affect the
measured acceleration variance as the fit interval is varied.
As fit interval is decreased,
the accelerations measured for particular trajectories will increase
because there is less coarse graining over turbulent fluctuations
and noise.
In addition, as the fit interval is decreased
the sampling of trajectories becomes inclusive
because the particles need not remain in view for as long an interval.

To determine how the extrapolation to zero fit time is related
to the true acceleration variance, we used a simulation
of the detection process.
The simulation is based on
three dimensional particle trajectory  data from Vedula and Yeung's
DNS simulation \citep{vedula:1998} at $R_\lambda = 240$,
which is known to have $a_0 = 3.44$.
These trajectories were
dimensionalized using the viscosity of water and the rate of energy
dissipation which would be needed to produce the same
Reynolds number in our apparatus.
These scaled trajectories were used as input to a computer model
which simulated the illumination of the particles
and the imaging of our measurement
volume onto the strip detectors, including diffraction and defocusing
of the imaging system.
The computer model also
simulated charge collection by the strip detector, including
correlated and uncorrelated noise, charge diffusion, and inoperative pixels.
The intensities produced by the simulation were then run through
the real-time thresholding and compression algorithms used in
the experiment, and were subsequently processed by the same data processing
algorithms used for actual experimental data.

\begin{figure}
\begin{center}
\epsfxsize= \sfw
\mbox{%
\epsffile{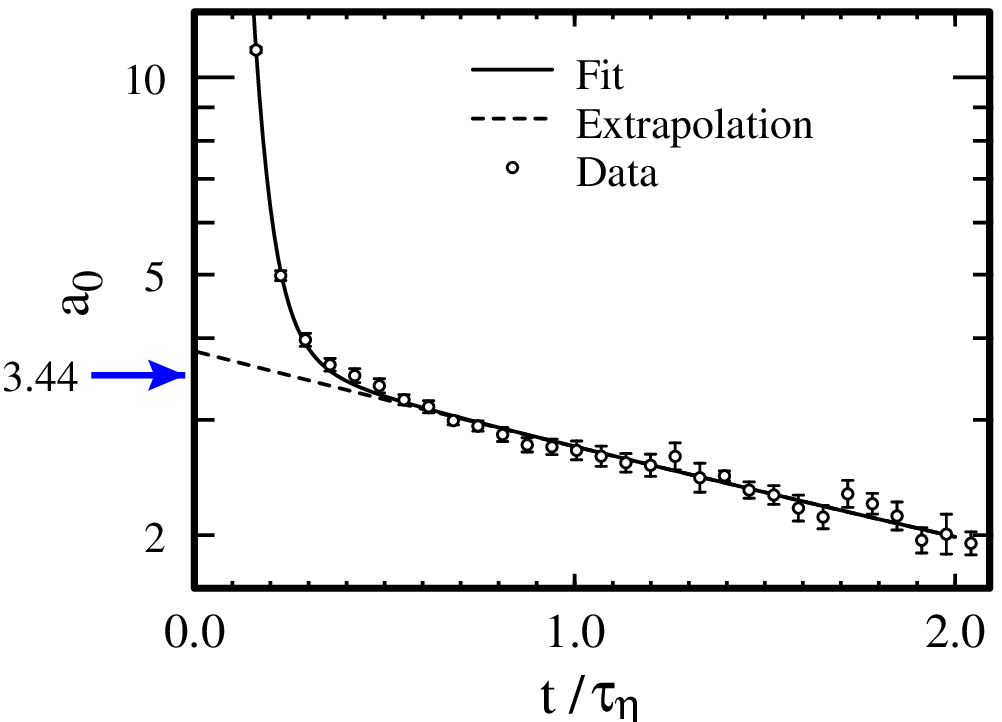}%
}
\end{center}
\caption{The plot shows $a_0$ as a function of normalized fit time
$\tau_f/\tau_\eta$ for trajectories obtained from direct numerical
simulation of turbulence\protect \citep{vedula:1998}.
The value of $a_0$ obtained from direct calculation from the simulation is 3.44.
The extrapolation technique overestimates this value by approximately 10\%.}
\label{fig:a0sim}
\end{figure}

The measurement of $a_0$ as a function of normalized fit interval
is shown in figure~\ref{fig:a0sim} (which may be compared with
figure~\ref{fig:a0_fit}, above).
The exponential dependence on $\tau_f$ is reproduced for large
$\tau_f$, as is the power law dependence for small $\tau_f$.
The extrapolation for $\tau_f \rightarrow 0$ gives a value
of 3.80.
This value is 10\% higher than the value of 3.44 which was
measured directly from the simulation \citep{vedula:1998}.
All normalized acceleration variance measurements presented in this paper
have been rescaled to correct for this overestimation.

\section{Effect of Finite Particle Size on Particle Accelerations}
\label{sec:psize}

An important question which must be considered is the extent to
which the polystyrene tracer particles are equivalent to ideal
fluid particles. In order to address this question, we have
repeated our acceleration measurements using tracer particles with
a range of particle diameters and fluid densities.  Not only are
these measurements essential to validate the Lagrangian
measurements, but the data for large particle sizes offers a new
perspective on the motion of finite size particles in turbulence.

Phenomenologically, we can identify two mechanisms by which a
discrepancy between statistics of tracer particle trajectories and
fluid particle trajectories could arise. (1)  It is possible that
the tracer particle's density is different from that of the fluid,
so that it experiences an acceleration that is different than the
acceleration that would be experienced by a fluid particle in the
same location. We call this the density dependent effect.  (2) It
is possible that the tracer particle's size is large enough that
the unperturbed flow would change significantly over the volume of
the particle.  Since this effect remains even when the tracers are
density matched, we call this the density independent effect.  The
dimensionless numbers that measure these effects are the ratio of
the particle to fluid density, $\rho_p/\rho_f$, and the ratio of
the particle diameter to the smallest length scale in the flow,
$d/\eta$. A commonly used parameter is the ratio of the Stokes
time ($d^2 \rho_p/\nu \rho_f$) to the Kolmogorov time
($\eta^2/\nu$) which is the combination of the size and density
parameters. 

Particle motion in fluid flows has been extensively studied both
as a fundamental fluid dynamics
topic \citep{Basset:1888,Corrsin:1956,Maxey:1983:EMS} and in order to validate measurement
techniques that rely on tracer
particles \citep{Buchave:1979:MTL}.
Most of the work in this area has focused on the
dynamics of particles when the density independent effect is
negligible. Not only is this regime amenable to exact theoretical
modelling, but it also is a good approximation for the tracers used
in many particle image velocimetry and laser Doppler anemometry
experiments.  In these experiments, it is desirable to have the
spacing between particles less than or equal to the size of the
smallest flow structures, so the particle size must be much
smaller than the smallest structures.

In order to determine the effects of finite particle size on the
acceleration measurements, we repeated the measurements at $\Rl =
970$ using particles of diameters between 26~$\mu$m and 450~$\mu$m
( 1.44~$\eta$ and 26~$\eta$).  The Stokes times for these
particles ranges from 0.67 ms to 216 ms, which should be compared
to the Kolmogorov time of 0.32 ms. We also measured acceleration
of 450~$\mu$m particles in NaCl solutions with densities between
1.00~${\mathrm g}\,{\mathrm cm}^{-3}$ and 1.11~${\mathrm
g}\,{\mathrm cm}^{-3}$. Since the particle density is
1.06~${\mathrm g}\,{\mathrm cm}^{-3}$, this range of densities
encompasses both negative and positive particle buoyancy.

\begin{figure}
\begin{center}
\epsfxsize=\smallfigurewidth
\mbox{%
\epsffile{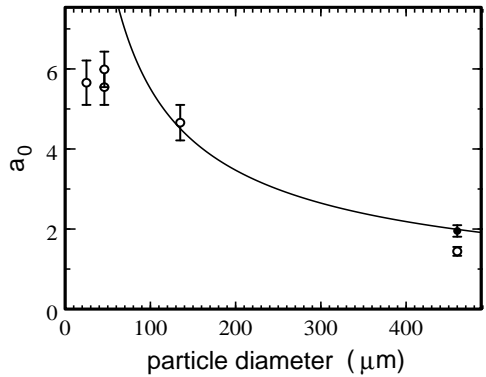}%
}
\end{center}
\caption{Normalized acceleration variance ($a_0$) as a function of
particle diameter 
 at $\Rl = 970$.
Open circles with error bars are  for particles of relative density 
1.06.
The closed circle with error bars is  for particles of relative density 
1.0.
The Kolmogorov length is $18\, \mu \mathrm{m}$ 
and the relative density is defined as $\rhoeff = \rho_{\mathrm
particle} / \rho_{\mathrm fluid}$.
The solid line shows $d^{-2/3}$ scaling.
} \label{fig:a0diameter}
\end{figure}

In figure~\ref{fig:a0diameter}, the measured Kolmogorov constant
$a_0$ is shown as a function of particle size for acceleration
data taken with pure water at $\Rl = 970$. A 70\% decrease is seen
in the measured acceleration variance between the smallest and
largest particles used.

\begin{figure}
\begin{center}
\epsfxsize=\smallfigurewidth
\mbox{%
\epsffile{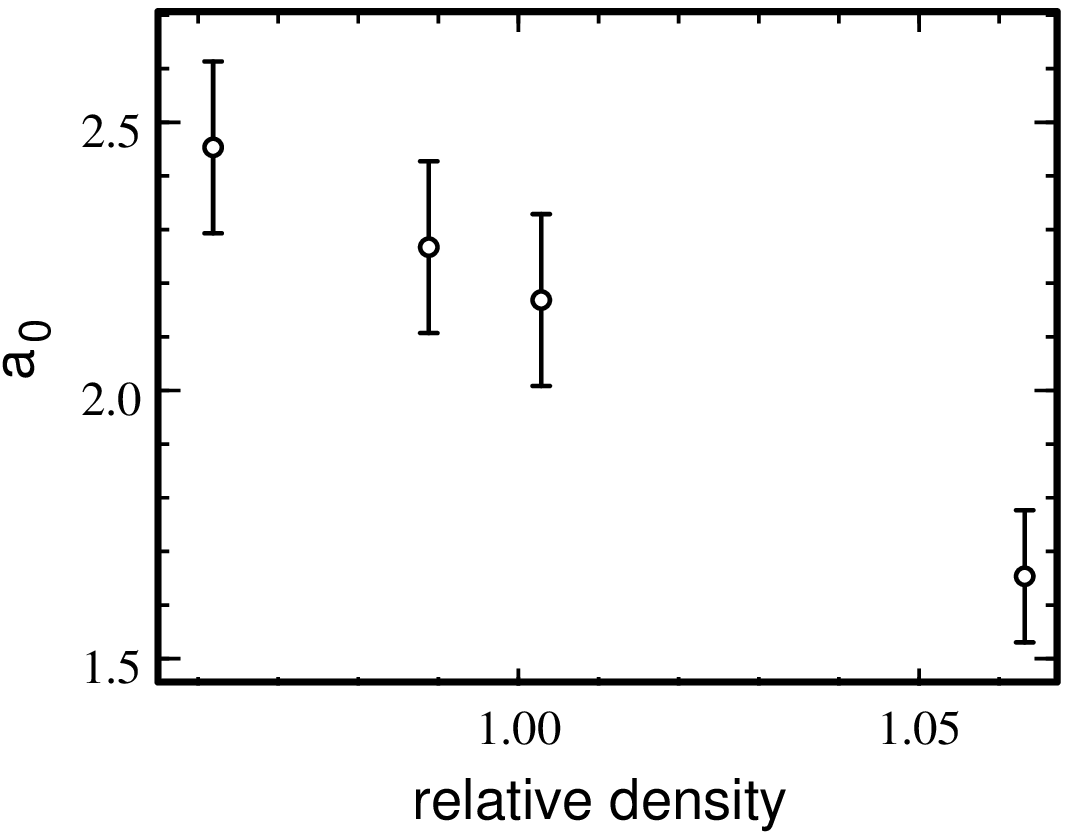}%
}
\end{center}
\caption{$a_0$ (normalized acceleration variance) as a function of
relative particle density for 450~$\mu$m particles at $\Rl = 970$.
} \label{fig:a0density}
\end{figure}

\begin{figure}
\begin{center}
\epsfxsize=\smallfigurewidth
\mbox{%
\epsffile{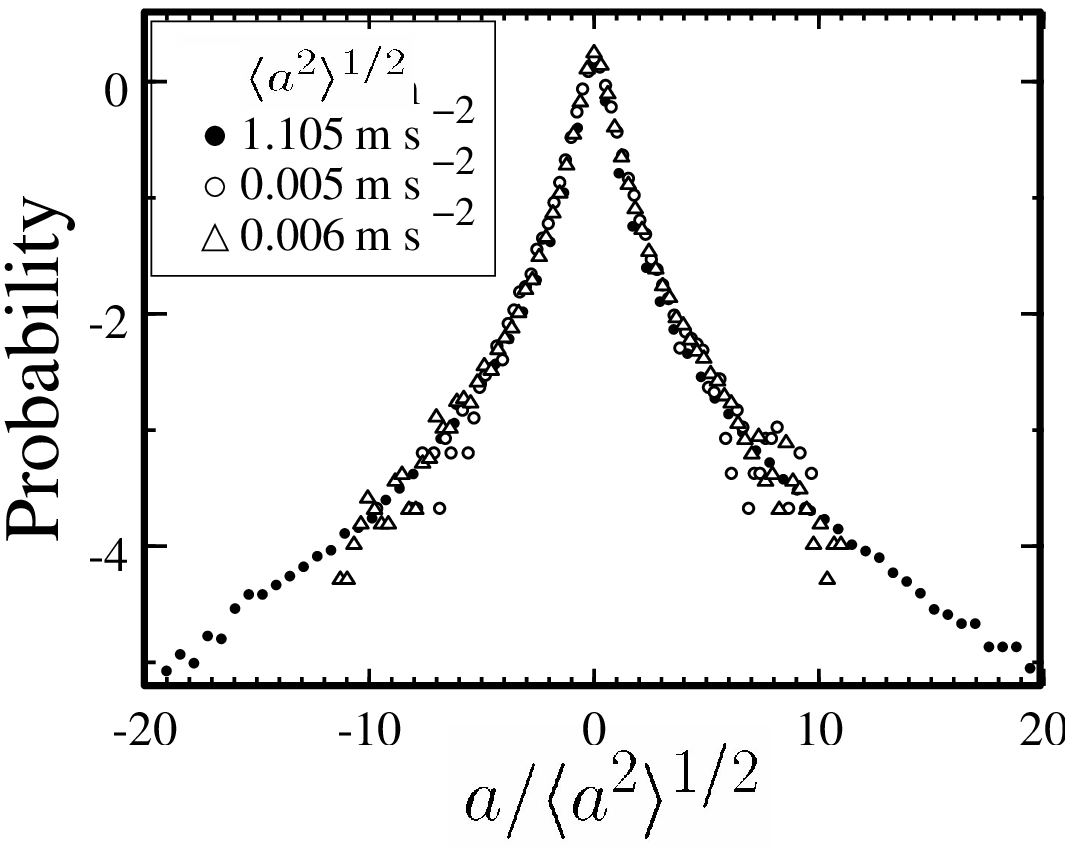}%
}
\end{center}
\caption{PDF of the acceleration for different particle sizes and fluid
densities at $R_\lambda = 970$ normalized by the rms acceleration. 
Closed symbols are for particles of size 50 $\mu\mathrm{m}$ and 
open symbols are for particles of size 450 $\mu\mathrm{m}$. 
Circles are for $\rho_{\mathrm{eff}} = 1.06$.
Triangles  are for $\rho_{\mathrm{eff}} = 0.96$.
The unnormalized standard deviation for each curve is  shown in the graph.
Out to 10 standard deviations the shape of the curve is not strongly affected.
} \label{fig:size_pdf}
\end{figure}

The dependence of the particle acceleration on the density
mismatch is shown in figure~\ref{fig:a0density}, where $a_0$ is
plotted as a function of relative density for 450~$\mu$m diameter
particles at $\Rl = 970$. In this case, the density of the fluid
is varied by the addition of NaCl, and the values of $a_0$ are
corrected for the slight change in kinematic viscosity which
occurs as the density is varied. The data for pure water
appears at effective particle density
$\rhoeff = 1.06$, and reflects the strong underestimate of
particle accelerations with large dense particles. If the
suppression of $a_0$ were due to density mismatch, the effect
would reverse in the case where the particles are lighter than the
fluid they displace, and the $a_0$ values would exceed the small
particle limit. The data at $\rhoeff = 0.96$ in
figure~\ref{fig:a0density} (taken at high salt concentration)
shows that although $a_0$ increases somewhat, it still remains far
below the small-particle limit of 6 as seen in figure~\ref{fig:a0diameter}.   
Figure~\ref{fig:size_pdf} shows the PDF of the acceleration for different
particle sizes and fluid densities. Out to 10 standard
deviations,  the shape of the distribution is not
strongly affected by particle size. 
This data demonstrates that,
in our experiment, the effective coarse graining of the acceleration by the finite
size particles is more important than the differential buoyancy
due to the density mismatch in our measurements.

A simple model based on the K41 phenomenology can be used to
predict the scaling of the acceleration with particle size in this
regime where density mismatch is not important.   We assume that
the effect of finite particle size is that only flow structures
larger than the particle size contribute to the acceleration. 
This means that the scaling of the acceleration variance can be
determined by starting with, $<a_x^2> \sim \epsilon^{3/2}
\nu^{-1/2}$, and replacing the viscosity with the value that would
make the Kolmogorov scale equal to the particle diameter, $\nu~
\rightarrow~\epsilon^{1/3} d^{4/3}$.  Consequently, the
acceleration variance should scale as $d^{-2/3}$  for large particle sizes.
We expect that below some
particle size this scaling is no longer adhered to, and 
the acceleration becomes independent of particle
size. 
 The solid line
in figure~\ref{fig:a0diameter} shows the $d^{-2/3}$ scaling.
There is not enough data to confirm this theory, but the data is
consistent with this scaling.   For large particle size, the data
can be interpreted as following the $d^{-2/3}$ scaling, and for
particle size less than 100~$\mu$m (5 $\eta$) there is a turn over
to the acceleration being independent of particle size.

To provide an upper limit on the possible deviation of tracer
particle acceleration from that of fluid particles, we note that
the dependence of $a_0$ on particle diameter in
figure~\ref{fig:a0diameter} could be interpreted as being roughly
linear. An extrapolation of the linear dependence to zero particle
size indicates that the acceleration variance of the 46~$\mu$m
particles used for the acceleration measurements reported above is
within 7\% of that of fluid particles. This is really a worst case
scenario, since the values obtained for 26~$\mu$m and 46~$\mu$m
particles are indistinguishable within experimental uncertainties,
and it is expected that the particle size dependence has zero
slope at the origin.   Note that this data is at the largest
Reynolds number reported in this paper, and the particle size
dependence will be weaker at lower Reynolds numbers. We conclude
that the 46~$\mu$m particles are small enough to measure the
acceleration variance at the 10\% accuracy which we have reported.

\section{Conclusion}

We have reported high resolution measurements of the Lagrangian
properties of a high Reynolds number turbulent flow. The particle
tracking system, based on the CLEO~III silicon strip detector, is
able to measure positions with spatial resolution of 0.5~$\mu$m
(1/40 of the Kolmogorov distance, or 1/6400 of the field of view)
and temporal resolution of 14~$\mu$s (1/20 of the Kolmogorov time)
in an $R_\lambda = 970$ flow.

The primary results of this paper concern the fluid particle
accelerations which, unlike the particle velocity, cannot be
measured with an Eulerian (fixed) probe.  We tested the
long-standing Heisenberg-Yaglom prediction that the acceleration
variance scales as $\epsilon^{3/2}$, (or $\tilde{u}^{9/2}$). At
low Reynolds number, our data is consistent with DNS. 
However, we find that for $500 \ge R_\lambda
\ge 970$ the value of $a_0$ is approximately constant, indicating
that K41 theory approximates the acceleration
scaling in this range of $R_\lambda$ well. Turbulent intermittency
models generally predict a very weak dependence of $a_0$ on
Reynolds number (for example, $R_\lambda^{0.135}$ in Borgas'
model) but experimental errors currently make it impossible to test
such predictions.

The fact that the flow between counter-rotating disks has a
well-defined anisotropy allows us to test the coupling of the
acceleration to the structure of the large scale flow. While the
velocity exhibits a significant anisotropy which is independent of
Reynolds number, the acceleration exhibits a smaller anisotropy
which decreases as the Reynolds number is increased, as shown in
figure~\ref{fig:a0_ratio}.

The autocorrelation function of the acceleration component was
also measured.  Because of measurement volume effects, this
measurement was restricted to relatively high Reynolds numbers.
The functions obtained agree with low $\Rl$ DNS simulations,
crossing zero at approximately 2.1~$\tau_\eta$, and exhibit
Kolmogorov scaling of the time.

The acceleration is found to be an extremely intermittent
variable, as evidenced by the very long stretched exponential
tails of the acceleration component PDF, shown in
figures~\ref{fig:adist} and \ref{fig:adist_7xy}. The level of
intermittency, reflected in the acceleration component flatness,
is found to increase with Reynolds number.  This is consistent
both with the general expectation that turbulent intermittency
should increase as the width of the scaling range increases, and
with DNS measurements of the flatness at low Reynolds number.

The fast temporal response of the particle tracking system allows
us to resolve the extremely violent events that make up the tails
of the acceleration component PDF. The trajectory in
figure~\ref{fig:traj} represents one such event, in which the
particle acceleration rises to
16,000~$\mathrm{m}\,\mathrm{s}^{-2}$, 1,600 times the acceleration
of gravity, and 40 times the rms value.  Although we do not have
quantitative evidence, we make the observation that the high
acceleration events appearing in the tails of the acceleration
distributions (such as figure~\ref{fig:traj}) seem to be
associated with coherent structures which persist for many
Kolmogorov times, substantially longer than the correlation time
of the acceleration components.

Particular attention has been paid to the determination of
measurement errors. We have developed a numerical simulation of
the detection process that takes input trajectories and models the optics
and electronics to create artificial data. Using this approach we have measured
the extrapolation error in the acceleration variance data and corrected for it.
We have also confirmed that our analysis codes and choices for the analysis 
parameters are not biasing the results.

The dependence of the particle acceleration variance on the tracer
particle size and density was also measured. It is found that the
acceleration observed with 46~$\mu$m diameter polystyrene
particles is within 7\% of the value that would be obtained from
ideal fluid particles, even at the highest Reynolds numbers
studied. We find that particles of diameter 450~$\mu$m (26~$\eta$)
have an acceleration variance that is a factor of 3.6 smaller than
fluid particles. In our experiment, this is caused primarily by the size of the
particles and is only slightly affected by their density mismatch
with the fluid. These measurements validate our Lagrangian
acceleration measurements, and in addition they highlight the need
for a deeper understanding of the motion of large density matched
particles in turbulence.

The techniques and measurements presented in this paper suggest
many possibilities for the future development of optical particle
tracking in turbulence.  The use of high resolution imaging
equipment and careful attention to measurement errors will
continue to be essential in future measurements. Possible
extensions of the techniques presented here include lengthening
the tracking times, improving the resolution of the
three-dimensional measurements and tracking larger numbers of
particles simultaneously.  These will allow precise high Reynolds
number experimental measurements of additional quantities
including relative dispersion, scaling in the inertial time range,
and the geometry of multi-particle Lagrangian motion.

\bf Acknowledgements \rm

We would like to thank the National Science Foundation for generous support under grant number
PHY9988755. We are grateful to R. Hill, M. Nelkin, S. B. Pope, E. Siggia,  Z. Warhaft
and P. K. Yeung for  stimulating discussions and suggestions throughout the project.
Thank you to R. Hill for carefully reading the manuscript. 
We  thank P. Vedula and P. K. Yeung for providing particle trajectories from their simulation data
at $R_\lambda = 240$ from \cite{vedula:1998} that were used to simulate the
detection process as described in Section~\ref{sec:simulation}. We also thank  C. Ward for
helping with the initial development of the detector.

\appendix

\section{Extraction of Particle Tracks from Intensity Data}
\label{sec:trackextraction}

The raw data from the strip detector consists of a series of
\td intensity profiles taken at regular time intervals.
The task of recognizing peaks in these intensity profiles and
assembling them into \ttd or \tttd tracks is, in principal, similar
to the case of \ttd CCD
images \citep{virant:1997,ott:2000,dracos:1996:TDV}.
However, the strip detector poses unique challenges.
Particle tracks cross much more often in \td than
in \ttd, and flaws in the detector lead to a number of
inoperative strips.
Both of these effects cause frequent drop-outs in tracks, which
would result in fragmentation of the trajectories.

The strategy for extracting tracks from the intensity data
is as follows.
The task is divided into four phases, peak detection,
track building, track splicing and track filtering.
In cases where \ttd trajectories are investigated, the tracks
are then passed to a matching algorithm which associates
$x$ and $y$ tracks to form \ttd trajectories.

\noindent
\textbf{Peak Detection.}
The algorithm scans each frame and identifies distinct groups of
above-threshold strips, and searches each group
for one or more  peaks, taking into account the
existence of inoperative strips.
Valid  peaks are passed to a routine which
calculates moments of the peak and fits it to a Gaussian
function, returning properties including peak centre, amplitude,
full-width at half maximum, area skewness and flatness.
The output of the peak detection algorithm is a list of peaks
for each frame.

\noindent
\textbf{Track Building.}
The track finding algorithm assembles the peaks found by the peak
detector into time-continuous tracks.
This algorithm builds tracks incrementally, extrapolating each track
forward one frame and searching for a peak which is the continuation
of the track.  If any ambiguity is found---if there is more than one peak
which might continue the track, or if there is more than one track which
might continue onto a peak---the track building algorithm fails to resolve
the match and begins new tracks as necessary.

\noindent
\textbf{Splicing.}
The list of tracks generated by the track builder is passed to a splicer which
connects track fragments which the track builder failed to connect
due to ambiguities.
Such ambiguities are quite common in \td projection, since tracks can
cross in \td projection even when the \ttd trajectories are distinct.
The track builder looks more than one frame ahead, and compares forward
extrapolations of the ends of tracks to backward extrapolations of the
beginnings of tracks.
It uses an iterative algorithm to make the best matches, while leaving
ambiguous matches unresolved.

\noindent
\textbf{Filtering.}
After the splicing algorithm has exited, the tracks are processed by
a filtering algorithm.
The algorithm will delete data points which meet certain criterion.
These criterion include upper and lower limits on the amplitude, and upper
limits on the width and flatness of the peak.
The algorithm also deletes data points which coincide with a list of
inoperative strips.
The idea is to keep marginally accurate data during the track
assembly process to maintain the continuity of the tracks,
but to exclude such data from subsequent statistical analysis, where
the large uncertainties could be detrimental to the analysis.

\noindent
\textbf{Coordinate Matching.}
Most of the acceleration data reported below is derived from
\ \ttd  trajectories.  To obtain \ttd trajectories, it is necessary to
match the $x$ vs.\ $t$ and $z$ vs.\ $t$ trajectories which are obtained from
the two strip detectors.
All such data is taken in the configuration shown in figure~\ref{fig:app}(a),
in which the same image is projected on two strip detectors, and only the
segments of tracks that lie in the overlap region of the two detectors
(See figure~\ref{fig:sdpair}) are used.
In this case, the optical intensity recorded by the two detectors for
a single track is highly correlated.
The autocorrelation function between the two trajectories is therefore
a good figure of merit for matching the tracks.

\section{Reinterpretation of previous data.}
\label{sec:old_data}

The data presented in this paper should be compared with
previously published measurements of the same quantities in the
same turbulent flow, but using a conventional position sensitive
photodiode instead of the strip detector to measure particle
trajectories \citep{Voth:1998:LAM}. The previous study reached the
correct conclusion that the Heisenberg-Yaglom scaling of the
acceleration variance is observed at high Reynolds numbers, but
because of limitations of the detector technology the numerical
values obtained for the physical quantities involved were
inaccurate.

The discrepancy can be largely attributed to the high noise level in
the position sensitive photodiode that resulted in much greater
uncertainty in the particle position measurements
and which necessitated the use of much larger tracer particles.
The intrinsic bandwidth of the UDT DLS-10 position sensitive
photodiode used in the previous experiments is approximately 5~MHz,
but temporal averaging of the
signal was used to increase the effective position resolution
at the expense of temporal resolution.
Signal averaging with an effective bandwidth of 100~kHz
was used in conjunction
with tracer particles of diameter 250~$\mu$m or 450~$\mu$m,
which gave a position resolution of 10~$\mu$m over a field of 1.5~mm,
corresponding to a dynamic range of 150.
By contrast, the strip detector
at 70~kHz has a position uncertainty of
approximately 0.1
strips on a field of 512 strips (0.8~$\mu$m over a field of
4~mm), giving a dynamic range of 5000,
which is a factor of 30 improvement.
Another critical advantage of the strip detector is that it is able
make these position measurements with 46~$\mu$m diameter particles.
As reported in Section~\ref{sec:psize} above, the 250~$\mu$m particles
which were required by the position sensitive photodiode do not
follow the flow at high Reynolds number and result
in significant underestimate of the acceleration.

The position sensitive photodiode is strictly limited to single
particle statistics, so that the energy dissipation could not be
measured in terms of velocity structure functions.
Instead, the acceleration autocorrelation function was compared
with DNS results to estimate the Kolmogorov time, and this value
was used to determine the rate of energy dissipation.
Presumably due to the fact that the 250~$\mu$m particles did not
adequately follow the flow, the estimate of the Kolmogorov
time was a factor 2.37 larger than the current value, and using
$\epsilon = (\nu/\epsilon)^{1/2}$, this gave an estimate of the
energy dissipation which is a factor 5.63 below the current value
at a given propeller speed.
The rms velocity $\tilde{u}$ was also underestimated by 10\%
in the old experiment, and using equation~\ref{eq:rlambda} the new
values obtained for $\Rl$ are smaller than the old by
a factor of 0.52, so the previously reported range of $\Rl$
(985--2021) would correspond to (512--1037) using the new calibration.
The ability to work at much smaller $\Rl$ (down to 140)
results from the use of smaller particles, which do not settle out
of the flow even at very slow propeller speeds.

The Kolmogorov constant $a_0$ is calculated from
\begin{equation}
a_0 = \langle a_i^2 \rangle \frac{\nu^{1/2}}{\epsilon^{3/2}}
\end{equation}
and is also affected by this inaccuracy in the estimate of $\epsilon$,
and would have increased the value of $a_0$ by a factor of 17.
However, the particle size effects and low spatial resolution caused
the acceleration variance to be underestimated by a large factor.
In the old study, an extrapolation to zero fit time was attempted,
but large position errors obscured the exponential dependence of
acceleration variance on fit time that is
seen in figure~\ref{fig:a0_fit}.
What was interpreted as
``measurement noise'' was in reality a blend of measurement error and
and short-time turbulence contributions.
The algorithm used to calculate the acceleration variance
gave a value which was effectively coarse grained
over a time interval which was believed to be
2.5~$\tau_\eta$ using the old calibration,
but which is actually 6~$\tau_\eta$ using the new
calibration.
As may be seen from figure~\ref{fig:a0_fit}, this results in a substantial
underestimate of the acceleration variance.
The acceleration variance was further reduced by the particle size
effect associated with the use of 250~$\mu$m particles.
The fact that the value of $a_0$ reported in the previous experiment
($7 \pm 3$)
is close to the value now reported ($\approx 6$)
is due to the fortuitous cancellation
of errors in the measurement of the dissipation rate and of the
acceleration variance.

\bibliography{turbulence2,turbulence,acc_paper,amc}
\bibliographystyle{jfm}

\end{document}